\documentclass[preprint,11pt]{elsarticle}

\usepackage{graphics}
\usepackage{graphicx}
\usepackage{verbatim}
\usepackage{acronym}
\usepackage{color}
\usepackage{comment}
\usepackage{subcaption}
\usepackage[T1]{fontenc}
\usepackage[english]{babel}
\usepackage{amsmath,amsfonts,amsthm}
\usepackage{amssymb}
\usepackage[linesnumbered,ruled]{algorithm2e} 
\usepackage{csquotes}
\usepackage{url}
\usepackage{paralist}
\usepackage{verbatimbox}
\usepackage{verbdef}
\usepackage{times}
\newacro{WGCNA}{Weighted Gene Co-Expression Network Analysis}
\newacro{TPR}{True Positive Rate}
\newacro{FPR}{False Positive Rate}
\newacro{SWA}{Swendensen Wang Algorithm}
\newacro{AUROC}{Area Under the Reciever Operator Curve}
\newacro{AUC}{Area Under the Curve}
\newacro{MVDA}{Multivariate Data Analysis}
\newacro{EPLS}{Exploratory Projection to Latent Structures}
\newacro{RMSE}{Root Mean Squared Error}
\newacro{SVD}{Singular Value Decomposition}
\newacro{PCA}{Principle Component Analysis}
\newacro{ANOVA}{ANalysis Of VAriance}
\newacro{mRNA}{messenger Ribonucleic Acid}
\newacro{RNA-seq}{Ribonucleic Acid sequencing}
\newacro{string-db}{STRING Database~\footnote{https://string-db.org/}}
\newacro{MNIST}{Modified National Institute of Standards and Technology}
\newacro{GSE2508}{\emph{Expression profiling in adipocytes of obese humans}}
\newacro{DBSCAN}{Density-Based Algorithm for Discovering Clusters}
\newacro{UMAP}{Uniform Manifold Approximation and Projection}
\newacro{CCA}{Connectivity Clustering Algorithms}
\newacro{LCA}{Link Clustering Algorithms}
\newacro{AHC}{Agglomerative Hierarchical Clustering}
\newacro{GRS}{Golden Ratio Search}
\newacro{GSS}{Golden Section Search}
\newacro{RNA}{Ribonucleic Acid}
\newacro{UMAP}{Uniform Manifold Approximation and Projection}
\usepackage{listings}
\usepackage{hyperref}
\usepackage{pgfplots}
\usepackage{tikz}
\usetikzlibrary{shapes,arrows,automata,positioning}

\pgfmathdeclarefunction{gauss}{2}{
  \pgfmathparse{ (\yoffset+1/(sqrt(2*pi*#2))*exp(-((x-#1)^2)/(2*#2)) )}
}
\usetikzlibrary{intersections}
\usetikzlibrary{decorations.text,decorations.pathmorphing}
\pgfplotsset{every axis plot post/.append style =
  {samples=80, smooth, thick,  mark=none} }
\usepackage{jigsaw}
\makeatletter
\def\ps@pprintTitle{%
  \let\@oddhead\@empty
  \let\@evenhead\@empty
  \def\@oddfoot{\reset@font\hfil\thepage\hfil}
  \let\@evenfoot\@oddfoot
}
\makeatother
\begin{document}
\begin{frontmatter}
%
\title{ Happy and Immersive Clustering Segmentations of Biological Co-Expression Patterns }
%
\author[kth,sfl]{Richard Tj\"{o}rnhammar}
\ead{richardt@kth.se}
\address[kth]{KTH Royal Institute of Technology, SE-100 44 Stockholm, Sweden}
\address[sfl]{SciLifeLab, Tomtebodavägen 23, SE-171 65 Solna, Sweden}

\begin{abstract}

In this work, we present an approach for evaluating segmentation strategies and solving the biological problem of creating robust interpretable maps 
of biological data by employing wards agglomerative hierarchical clustering applied to coexpression coordinates to deduce a faithful representation 
of the input.

We adopt and quantify two analyte-centric metrics named happiness and immersiveness, one for describing the suitability of a single analyte concerning 
the segmentation as well as a second metric for describing how well the segmentation catches the underlying data variation. We show that these two 
functions drive aggregation and segregation of segmentation respectively and can produce trustworthy segmentation solutions. We discover that the 
immersiveness metric exhibits higher-order phase transition properties in its derivative to cluster numbers.

Finally, we find that the cluster representations and label annotations, in the case with clusters of high immersiveness, correspond to 
compositionally inferred labels with the highest specificity. The interconnectedness mirrors the potential relationships between cluster representations,
label annotations, and inferred labels, emphasizing the intricate nature of biology and the representation of the specific expressions of 
gene products.

\end{abstract}
%
%
\end{frontmatter}

\section*{ Introduction }

\subsection*{ Background }

One key feature of modern biotechnology platforms is the excessive amount of viable information present for a specific sample. In the fields of 
proteomics and transcriptomics, genetic products are quantified to assess the emergent properties of the studied systems. This can include both 
protein-coding as well as non-protein-coding gene products, where the former has traditionally been the main interest of many research 
endeavours~\cite{uhlen_tissue-based_2015,uhlen_pathology_2017}.

A large number of measured analytes, used to describe a single system, effectively means that the problem is overdetermined. Many techniques 
exists and are used to conduct dimensionality reduction while conserving the amount of useful information for describing the 
system state~\cite{becht_UMAP_2019,narayan_assessing_2021,TDA_annurev,KeplerMapper_JOSS,mcinnes_umap_2018,golub_svd_1970}.
These include a large array of, what today is thought of as machine learning, techniques.  Some of the most widely used in 
bioinformatics that relates to our approach include, but are not limited to clustering techniques~\cite{10.5555/3001460.3001507,CFinder_Palla_2005}, 
regression models or factorisation-based models~\cite{girolami_mercer_2002} as well as classifiers such random forest models~\cite{MLtine2023,Xiong2012RandomFF} or 
can make use of batch correction approaches alongside clustering~\cite{jointly2023}.

Our distance dependant formalism is thereby highly
reminiscent of co-expression clustering~\cite{lemoine_gwena_2021} and other \ac{PCA} based methods~\cite{zhang_fast_2024}.
We are not deducing a scale-free representation~\cite{ScaleFreeNetwork-Broido_2019} in a directed fashion
of the clustering solution such as is potentially done in \ac{WGCNA} methods~\cite{WGCNA_ZaH_2005,schlosser_netboost_2021}. The main reason is that the
hierarchical clustering approach, in general, cannot be assumed to describe free-scaling network modules.

In order not to calculate the pseudo-inverse of the expressions and use it to solve the biological problem for the sample encodings, we adopt a 
factorisation scheme. By diagonalisation and utilizing the insight that the factorised matrices are globally aligned component matrices of the full data
we may directly attribute features with sample group properties. This can be realised by studying the SVD~\cite{golub_svd_1970} 
method and realizing that a mean grouping of component factors in sample space corresponds to the mean grouped sample value before diagonalisation 
and that they are aligned with the feature components post-diagonalisation. From the decomposition, one can also deduce an adequate distance matrix 
representation that can be used for clustering.

Clustering is a rich field with many popular approaches that also include neighbor-based approaches such as the ones developed by 
Seurat~\cite{kiselev_sc3_2017}, Cluster Finder~\cite{CFinder_Palla_2005} and Louvain or Leiden~\cite{traag_louvain_2019} clustering. Its long history 
stems from the need to coarse grain data and uncover hidden groups and data variation within a data set~\cite{girolami_mercer_2002}. Unsurprisingly, 
clustering has not only a 
long history but also a multifaceted epistemological heritage with ties to many disciplines with applications in biological sciences using physical 
approaches~\cite{Murua2008,agrawal_potts_2003}. Most clustering approaches are employed to describe nonlinear effects and variations in data and thereby 
share a fundamental relationship with nonlinear \ac{PCA}~\cite{scholkopf_nonlinear_1998}, and while being well known is still being actively 
studied~\cite{tirelli_unsupervised_2022}.

In this work, we adopt Wards \ac{AHC}~\cite{ward_hierarchical_1963} when working with real-world data. We have chosen Wards method since it is believed to 
produce segmentations with minimal spatial distortion and minimized 
inter-cluster variances~\cite{fernandez_versatile_2020} to explore the coarse-graining properties of the component distance hierarchies.

\subsection*{ Clustering }

For a general clustering problem, we want to deduce a segmentation strategy of $K$ analytes into $M$ cluster label segments.

To make a connection between system fragmentation and system information in a physical sense for clustering we define a system 
constituent state, $x_i$ to have the probability $p_i$ with the following relation
\begin{equation}
\sum_i p_i = 1
\label{eq:onesum}
\end{equation}
so that we can form averages as
\begin{equation}
f_{avg}(\mathbf{x})\quad=\quad< f >\quad=\quad\sum_i p_i f( x_i )
\label{eq:probav}
\end{equation}
and define 
\begin{equation}
\mathcal{Z}(\beta)\equiv\sum_i e^{-\beta f(x_i)}
\label{eq:fparted}
\end{equation}
as the normalisation criterion when the system is subject to an inverse fractionation probability $\beta\propto T^{-1}$ in 
equilibrium~\cite{ghavasieh_diversity_2024} and it is clear that this choice maximizes the
entropy of system segmentation defined in the information space volume $\mathcal{I}$~\cite{plischke1994equilibrium}. We recognize the stationary system 
information energy score as
\begin{equation}
E = -\Big( \frac{\partial}{\partial\beta} \log{ \mathcal{Z}(\beta) } \Big)_{\mathcal{I}}
\label{eq:informationenergy}
\end{equation}
thereby via 
\begin{equation}
C_\mathcal{I} \equiv \Big( \frac{\partial E}{\partial T} \Big)_{\mathcal{I}}
\label{eq:CI}
\end{equation}
and using the fact that $\frac{\partial}{\partial T}=-\beta^2\frac{\partial}{\partial\beta}$ yields
\begin{eqnarray}
C_\mathcal{I} &=& \beta^2 \frac{\partial^2}{\partial\beta^2} \log{ \mathcal{Z}(\beta) } \\
&=& \beta^2 \big( < f^2 > - < f >^2 \big) \\
&=& \beta^2 \sigma^2_f
\label{eq:fluctuationrel}
\end{eqnarray}
which is recognizable as an information capacity for $f$ times the fractionation contribution so that we can define
\begin{equation}
C_f = \sigma^2_f
\label{eq:infocap}
\end{equation}

\begin{figure}[t!]
\begin{centering}
\includegraphics[width=15cm]{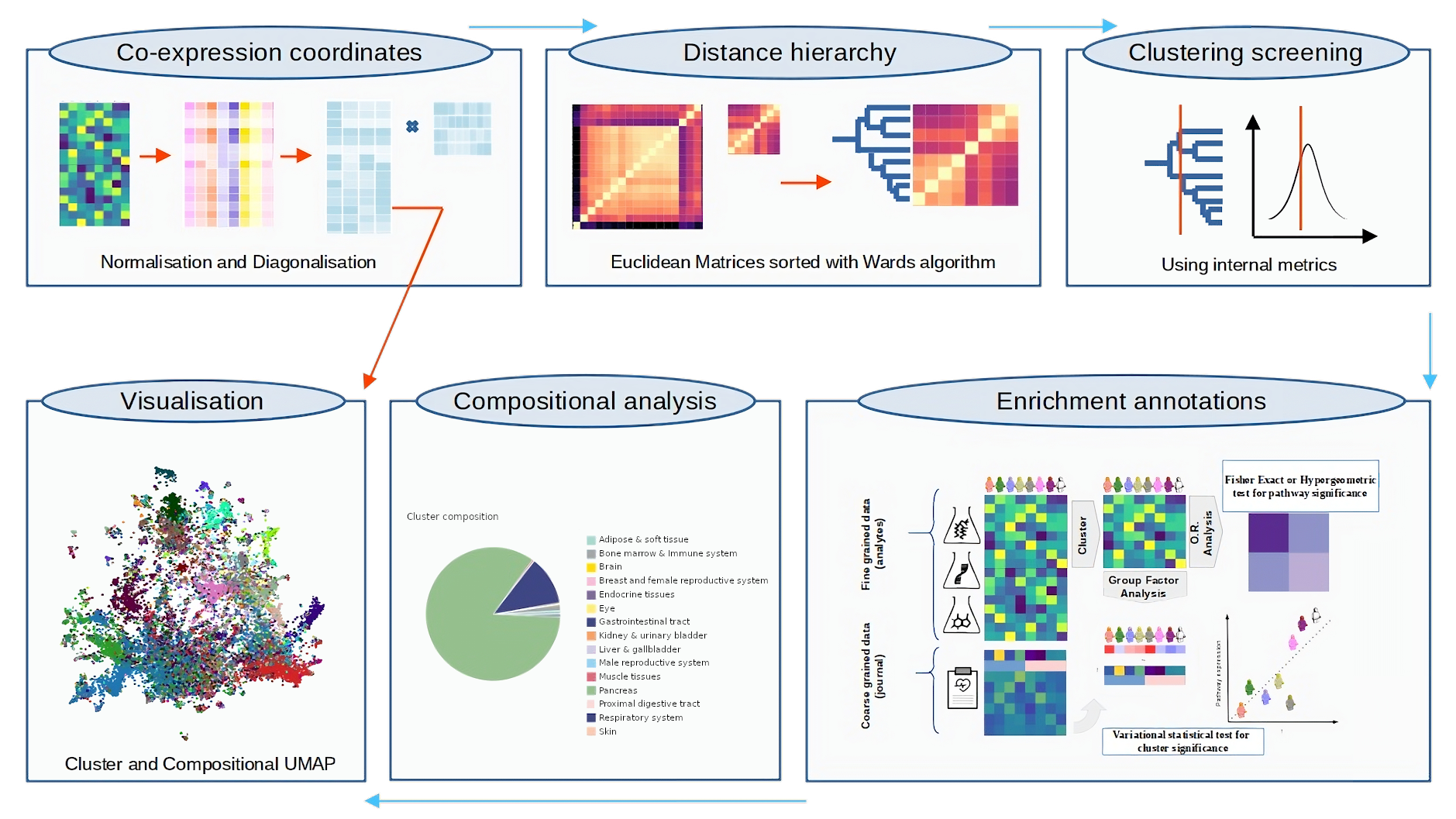}
\caption{ The publically available Biocartpgraph pipeline aims at delivering annotated clustering solutions of biological analyte data. This work is
focusing on the screening evaluation section of the biocartographer }
\label{fig:biocartograph}
\end{centering}
\end{figure}

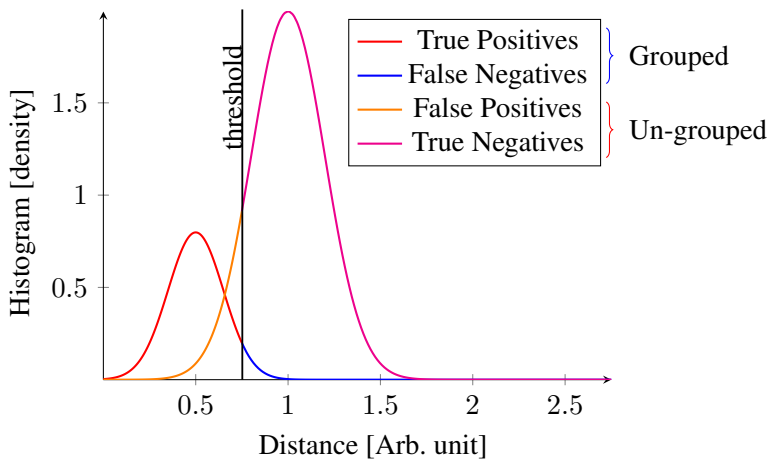
\begin{figure}
\begin{centering}
\resizebox{300pt}{!} {

\begin{tikzpicture}

  \draw[white, very thin] (-1.5,-1.5) rectangle (8,6);
  \draw[decorate, decoration = {text along path ,
    text = { Distance [Arb. unit] }}]
    (2.0,-1) -- (6.5,-1);

  \draw[decorate, decoration = {text along path ,
    text = { Histogram [density] }}]
    (-1.0,0.75) -- (-1.0,4.5);

  \begin{axis}[axis lines = center, axis equal image,
      domain = 0.0:2.75, 
      legend entries = { True Positives , False Negatives, False Positives, True Negatives } ]
    \addplot[red,  name path=TP, domain=0:0.75] { e^(-((x-0.5)/0.15)^2/2 )/sqrt(2*pi)/0.5 };
    \addplot[blue,  name path=FN, domain=0.75:2.75] { e^(-((x-0.5)/0.15)^2/2 )/sqrt(2*pi)/0.5 };
    \addplot[orange, name path=FP, domain=0:0.75] {  e^(-((x-1.0)/0.2)^2/2 )/sqrt(2*pi)/0.2  };
    \addplot[magenta, name path=TN, domain=0.75:2.75] {  e^(-((x-1.0)/0.2)^2/2 )/sqrt(2*pi)/0.2  };
    \path[name path=xaxis] (axis cs:0.0,0) -- (axis cs:2.0,0);

  \end{axis} 

  \draw[ decorate,decoration={brace,mirror,raise=1pt},blue] (6.75,4.0) -- (6.75,4.75);
  \draw[decorate, decoration = {text along path ,
    text = { Grouped }},  font=\tiny\sffamily\bfseries ]
    (7.0,4.25) -- (9,4.25);
  
  \draw[ decorate,decoration={brace,mirror,raise=1pt},red]  (6.75,3) -- (6.75,3.75);
  \draw[decorate, decoration = {text along path ,
    text = { Un-grouped }},  font=\tiny\sffamily\bfseries ]
    (7.0,3.25) -- (9,3.25);

  \draw[ line width=0.25mm ] (1.875,0) -- (1.875,5.0);
  \draw[decorate, decoration = {text along path ,
    text = { threshold }}]
    (1.875,3) -- (1.875,5.0);
\end{tikzpicture}
}
\end{centering}
\caption{Schematic of unsupervised multilabel inference evaluation }
\label{distROC}
\end{figure}

We are assessing the success 
of the clustering segmentation strategy based on how well the cluster label inferences describe the underlying distance matrix, 
such as in~\cite{clusterROC}. An assumption in clustering \ac{AUROC} calculations is that the label inferences are related via a 
similarity measure, such as the covariational structure, of data. This is not necessarily true for an arbitrary segmentation strategy but is the 
standard assumption made in co-expression-based clustering. Here we employ distance measures instead of similarity and the assessment approach in ~\cite{clusterROC} 
takes the form of Figure~\ref{distROC}. The covariation as well as the clustering label inference problem can be 
modelled using several different approaches and we define the coexpression coordinates to be calculated from the \ac{PCA} scores using a Euclidean 
distance metric. The useful dimension of the \ac{PCA} coordinates correspond to the minimal extension of the input data ($\mathbf{X}$) 
dimensions so that the number of components becomes $n_C = \min{\dim{\mathbf{X}}} - 1$. Which, for our datasets, are determined by the number of samples 
and sufficient to model all the information present in the datasets~\cite{thibeault_low-rank_2024}.

For our expression data matrix $\mathbf{X}$, containing $K$ features and $L$ samples, we define the standardised values 
$\mathbf{Z} = \frac{\mathbf{X}-\mu_i}{\sigma_i}$ such that we assume $Z_{ij} \in \mathcal{N}(0,1)$, implying that $\mu_i,\sigma_i$ are 
the expression sample means and standard deviation values respectively. Following $\mathbf{Z}$ our diagonalised component matrices becomes
\begin{equation}
\mathbf{U}\mathbf{S}\mathbf{V}^{T} = \mathbf{Z}
\label{eq:diagonalisation}
\end{equation}

where $\mathbf{P}$ defines the absolute coordinates of the covariation matrix as in equation~(\ref{eq:covariationcoordinates})
\begin{equation}
\mathbf{P} = \mathbf{U}\mathbf{S}
\label{eq:covariationcoordinates}
\end{equation}

From this, it is apparent that the covariation matrix can be calculated as
\begin{equation}
\mathbf{C} = \frac{1}{N-1} \mathbf{P}\mathbf{P}^T
\label{eq:covariation}
\end{equation}
that we employ to calculate a covariation distance given by the Euclidean distance between covariation coordinates
\begin{equation}
D^2_{ij} = \sum_{k=0}^{n_C} ( \mathbf{P}_{ik} - \mathbf{P}_{jk} )^{2}
\label{eq:covariatiodistances}
\end{equation}
This distance matrix is then further employed in an \ac{AHC} step that we screen to deduce a segmentation cut for
the clustering representation.

Various dimensionality reduction techniques can be employed to coarse-grain the system and reduce the complexity of the studied 
problem~\cite{becht_UMAP_2019,pedregosa_scikit-learn_2011,mcinnes_umap_2018}. 
Here we employ \ac{UMAP} transformations to the co-expression coordinates before clustering real-world data. The clustering approach is chosen to 
be a hierarchical approach in that it creates well-connected segmentations. For any cut through the hierarchy, we can directly determine the compositions and \ac{AUROC} 
metrics. This implies that once a segmentation strategy has been chosen we can assess when the underlying variation is no longer markedly improved by 
additional clusters. This sets the number of cluster states in our model.

Since the chosen segmentation across analytes can be compositionally studied we may deduce what 
compositional component is in majority as well as its compositional specificity. This forms a 
foundation for compositionally auto-annotating the analyte feature.
In the case where sample label groups are good approximatives of the underlying covariation this annotation corresponds to a compositionally inferred 
attribution, with high \ac{AUROC} values. We have chosen to employ the Gini coefficient, in place of many other available compositional 
coefficients~\cite{composition_benchmark_2016,gorecki_kendalls_2017}, mainly because of its broad usage but also because it is believed to share an 
approximate relationship with the clustering-based \ac{AUC}~$\approx\frac{gini+1}{2}$~\cite{clusterROC}.

Now we present the two new validation metrics for the variationally linked segmentation strategies for the attribution of high-dimensional biological 
data. Throughout this text, the input data $\mathbf{X} $ corresponds to publicly available \ac{RNA-seq} expression data coming from any of five datasets 
that are explicitly stated together with figures belonging to them. 

\section{Method} \label{sec:method}

The author developed the software package Biocartograph~\footnote{https://github.com/rictjo/biocarta}~\cite{biocartograph}, see Figure~\ref{fig:biocartograph} for
evaluation and automatic annotation of clustering solutions of biological data. 

Since we employ \ac{AHC} on diagonalised expression data the critical assumption in all the methodologies is the conceptual dependence on the ability of
the coordinates to recreate a covariation distance matrix. This distance matrix is employed in 
an agglomerative hierarchical clustering step, where the full hierarchy is determined and screened to deduce a viable starting position for the 
clustering representation. We employ a screening function, similar to what was done in 
previous work~\cite{COPT}, but forsake the optimisation and use the formalism to fully evaluate the hierarchical agglomerative clustering solution space by offsetting the 
internal metrics of the analytes versus the clusters. We propose two new metrics (called happiness and immersiveness) for assessing per analyte 
relevances in the following two sections.

\subsection*{ Happiness }
For any analyte belonging to any cluster in the segmentation, we can deduce the cluster attribution of its $N_e$ closest nearest neighbours. We postulate 
that the happiness of an analyte is defined in regards to how homogenous the neighbour cluster attribution of those closest neighbours is. We define the 
attribution function as $\nu(i)$, which returns the cluster label of an analyte $i$. The neighbour function $\eta(i,j)$ defines the neighbour $j$ of 
analyte $i$. We also define the unit Dirac delta function as $\delta_{a,b}$. Then we can formulate the happiness of the analyte $i$ as:
\begin{equation}
h(i) = \frac{1}{N_e}\sum_{j=1}^{N_e} \delta_{\nu(i),\nu(\eta(i,j)) }
\label{eq:happy}
\end{equation}
The value of $N_e$ represents the closest environment and is chosen to be $20$ in this text. To conduct the calculation over all neighbours
the sum in Eq.~\ref{eq:happy} becomes the sum over all $j$ in $\delta_{\nu(i),\nu(j)}$. The number of neighbours is always assumed to 
include itself in this work. As a result, we immediately understand that the happiness metric can be seen as an interior volume function of the clustering
solution and will as a result minimize interfacial area and promote forming a large homogenous solution. The metric is complementary to the immersiveness and 
favours aggregation. The average happiness is defined as:
\begin{equation}
H = \frac{1}{K} \sum_{i=1}^{K} h(i)
\label{eq:happiness}
\end{equation}

\subsection*{ Immersiveness }
The clustering segmentation \ac{AUROC} is calculated in the same fashion as has been done in previous works~\cite{clusterROC,hmAUC1982}, but also extended 
into the immersion score by calculating an analogous \ac{AUC} per analyte towards all others in the distance matrix. For analyte $i$, the $i$:th row of the 
distance matrix is used to construct the scores instead of the entire matrix. The full clustering \ac{AUROC} of the entire solution is defined 
in terms of all analyte distances and calculated as in~\cite{hmAUC1982}. The only distinguishing property in the distance matrix is whether the distance 
pairs are grouped in a cluster or not. Thereby the immersion is defined as one analyte to all other similarity values but is similar to the full \ac{AUC}
 calculation. We denote this immersion score, or per analyte ($i$) score, non-rank corrected \ac{AUC}, $\iota(i)$. We can readily understand that any 
\ac{AUC} function promotes many cluster labels as the cluster-id factors approach the underlying covariational structure that the distance matrix is 
calculated from. Thereby forming a monotonically increasing function versus the number of clusters in the segmentation solution ($M$) and promoting 
segregation. We define the average immersion as the immersiveness of the solution as:
\begin{equation}
I = \frac{1}{K} \sum_{i=1}^{K} \iota(i)
\label{eq:immersiveness}
\end{equation}
The immersiveness can be understood as a restricted \ac{AUC} metric where the distance matrix row-to-row interactions are included as an 
average interaction.

Here we elaborate on the calculation of the $\iota(i)$. To describe the immersion $\iota(i)$ we make use of the finite unit Dirac delta 
function (~$\delta_{ij}=\delta_{\nu(i),\nu(j)}$~) 
expressed as an incidence matrix. We note that this matrix describes when any pair is grouped while $1-\delta$ describes all ungrouped pairs. Now we have
\begin{eqnarray}
\Delta^g_{ij} & = D^{grouped}_{i,j}	& = D_{ij} \delta_{ij}  \\
\Delta^u_{ij} & = D^{ungrouped}_{i,j}	& = D_{ij} ( 1-\delta_{ij} ) \\
\label{eq:iota}
\end{eqnarray}
as the distance decomposition for our multilabel clustering solution. To construct the evaluation of the case discrimination, as in Fig.~\ref{distROC}, 
we employ the standard heaviside function $\Theta(x)$. This results in an expression for the \ac{TPR}
\begin{eqnarray}
g(i,d) & = 1 + \sum_{j=1}^{K} \Theta( d - \Delta^g_{ij} ) \\
TPR(i,d) & = g(i,d) / \int_0^\infty g(i,x) dx
\label{eq:fpr}
\end{eqnarray}
as well as one for the \ac{FPR}
\begin{eqnarray}
u(i,d) & = 1 + \sum_{j=1}^{K} \Theta( d - \Delta^u_{ij} ) \\
FPR(i,d) & = u(i,d) / \int_0^\infty u(i,x) dx
\label{eq:fpr}
\end{eqnarray}

Since these complementary metrics are expressed for all $i$ and evaluated at the same distance we can form
\begin{equation}
\iota = \int_0^1 TPR(FPR)d(FPR)
\label{eq:iota}
\end{equation}
and note that the integrand ROC curve, in equation~(\ref{eq:iota}), is our sought information functional $f$ for use in 
determining the information capacity, see equation~(\ref{eq:infocap}).


Furthermore, we define that $H$ is the average happiness of all analytes, at a hierarchy level $\xi$, while the $I$ function is the average of 
the immersions of the 
segmentation analytes, $\partial H /\partial \xi \le 0 \quad \forall \xi$ and that $\partial I /\partial \xi \ge 0 \quad \forall \xi$ with $\xi \propto T$. 
We thereby employ a weak meaning of the terms monotonically increasing or decreasing when assuming that the above derivative relationships are true 
respectively. The average 
immersiveness or \ac{AUC} both favour segregation and a straightforward way of assessing the fitness of the clustering segmentation is via the \ac{AUC}.
%

A unified segmentation score can be defined as:
\begin{equation}
\hat{G}(\xi,p,q) = \frac{ \hat{H}(\xi)^p \hat{I}(\xi)^q } { \hat{H}(\xi)^p + \hat{I}(\xi)^q } \quad
\quad \forall \xi,p,q > 0
\label{eq:geom}
\end{equation}
Where $\hat{H}$ and $\hat{I}$ are the max-min range normalized happiness and immersiveness functions respectively. The un-normalized functions are not 
guaranteed to form a unimodal $\hat{G}$ function, why it can only serve as a screening function once the full hierarchy solution has been obtained. 
Since the \ac{AUC} computation is computationally expensive it also does not serve as a good optimisation goal function.

\subsection*{ Compositional coefficients }

In our context, the sample labels in the labelled data do not refer directly to the clustering inferred labels. It is important to note that compositional 
coefficients can be calculated for both. In the first case, compositional information from the samples can be attributed to cluster labels. In the latter, 
cluster attribution of sample labels is inferred because compositional variance is segmented on the cluster-ID. The same compositional metric can either 
refer to the specificity for each cluster versus sample attribution or the cluster attribution of each unique sample label. We need to distinguish between 
sample groupings and clustering labels in a compositional context. One pertains to the analyte attribution and the second to the sample attribution.

Aggregating samples on common labels and retaining the summed expressions, the mean expressions, or the normalized expression of those samples across 
labels results in markedly different fractional composition information for those labels. Here we make the further distinction between compositional 
coefficients calculated for different qualities of the groupings.

We perform the cluster composition calculation in three steps. The first is to aggregate from samples to labels via direct summation to obtain analyte 
expressions for the composition of sample labels. Secondly, we normalize the expressions for each analyte to make analyte contributions comparable.
This is achieved by sum-normalization which ensures that highly and lowly expressed analytes retain their expression profile across the sample labels 
and their profiles contribute equally. This is done before finally aggregating a typical cluster profile via summation of analytes onto cluster labels.

\section{Results} \label{sec:results}

\begin{figure*}
\includegraphics[width=15cm]{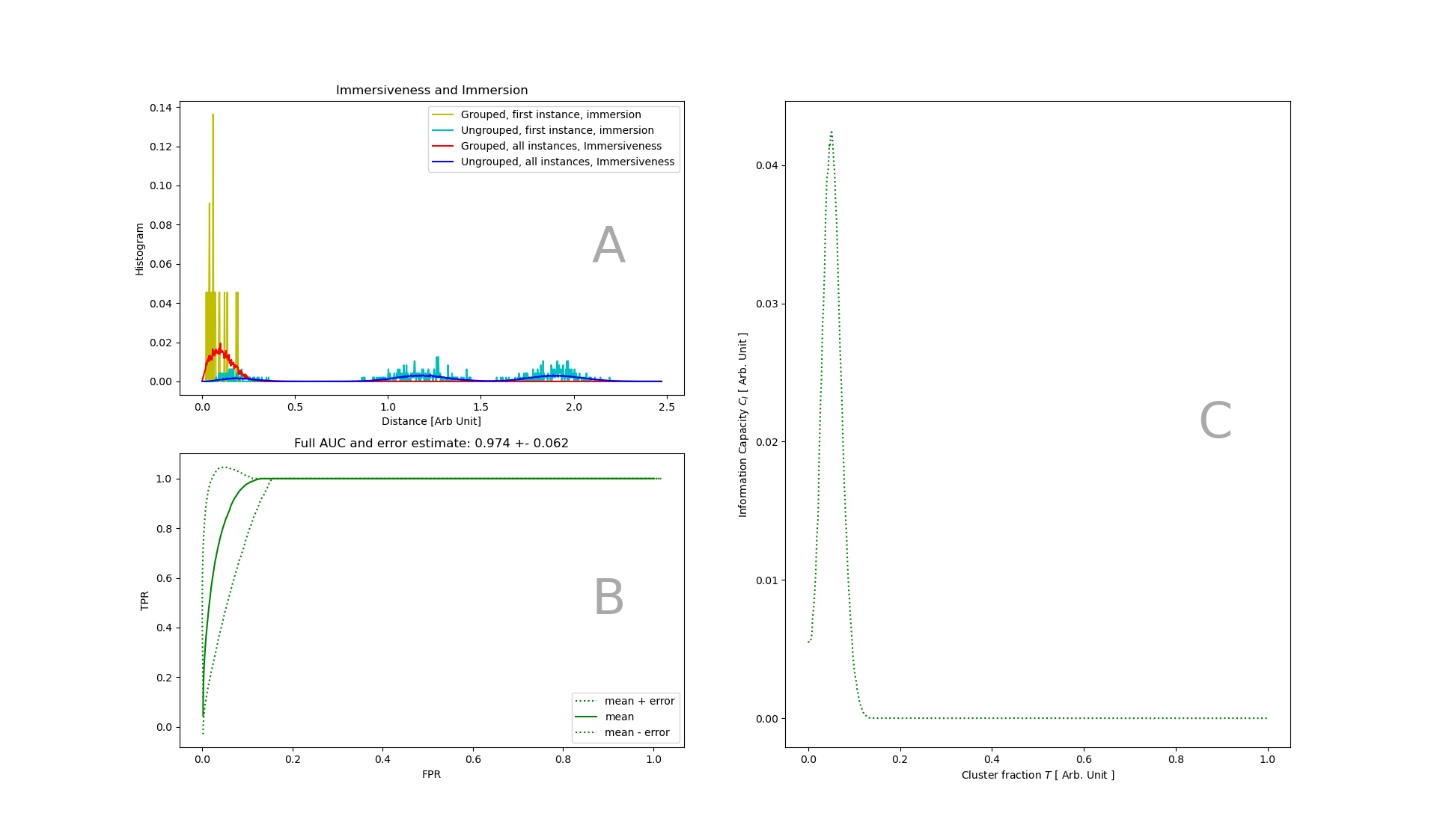}
\caption{Synthetic data example employing agglomerative hierarchical clustering and complete linkage (maximum distance) method for cluster distance attribution.
A~) Depicting a typical clustering solution immersiveness together with the first $\iota(i=1)$.
B~) From the standard deviation of the immersiveness from the $\iota$ we estimate the error bounds for the \ac{AUC}.
C~) The average $\iota$ fluctuations $\sigma_\iota$ can be interpreted as an information capacity across a domain where
$0$ translates to a single cluster and $1$ equates to a $K$ cluster solution. We call this domain measure, which measures the degree of
system fractionation, the temperature of the clustering solution. }
\label{fig:toycluster}
\end{figure*}

The overarching goal has been to find an unbiased rationale for representing the data well. We will study this in the context of hierarchical
clustering and the segmentations produced by it. All of the level cuts through all the hierarchical segmentation strategies for five different 
data sets were evaluated. A suitable representation of the
covariational structure was determined by studying both the information capacity as well as the unimodal cluster size-dependent function, in the form of 
equation~(\ref{eq:geom}). We will report on the
the interplay of those metrics, the phase transition point and the screening function as well as the compositional impact, in this section.

We first focus on the derived information capacity calculated in accord with equation~(\ref{eq:infocap}) in order to make an initial assessment of our 
modelling assumptions. 
This was conducted for several \ac{AHC} linkage approaches using simulated data, see Fig.~\ref{fig:toycluster}
for the complete linkage \ac{AHC} example. In Fig.~\ref{fig:toycluster}~A we depict the parts of a single immersion calculation as well as the full immersiveness 
decomposition. Our formalism allows us to directly calculate the standard deviation of both the TPR and FPR, which we can further employ to calculate
the error associated with each \ac{AUROC} value, depicted in Fig.~\ref{fig:toycluster}~B. The resulting standard deviation of the immersiveness or average
standard deviation of the immersions can via equation~(\ref{eq:infocap}) be interpreted as an information capacity that obtains max value during the 
information transition. This point coalesces with the maximum \ac{TPR} change subject to minimal change of the \ac{FPR}. From Fig.~\ref{fig:toycluster} it is clear
that our approach works for simulated data, see supplementary material for additional graphs and information. This approach was thereby further employed to 
study real-world datasets.

\def\mybool{0}
\def\halfwidth{175pt}
\def\fullwidth{350pt}
\begin{figure*}[t!]
  \centering
    \centering
  \if\mybool1
    \graphicspath{{img}}
    \def\svgwidth{0.475\textwidth}
    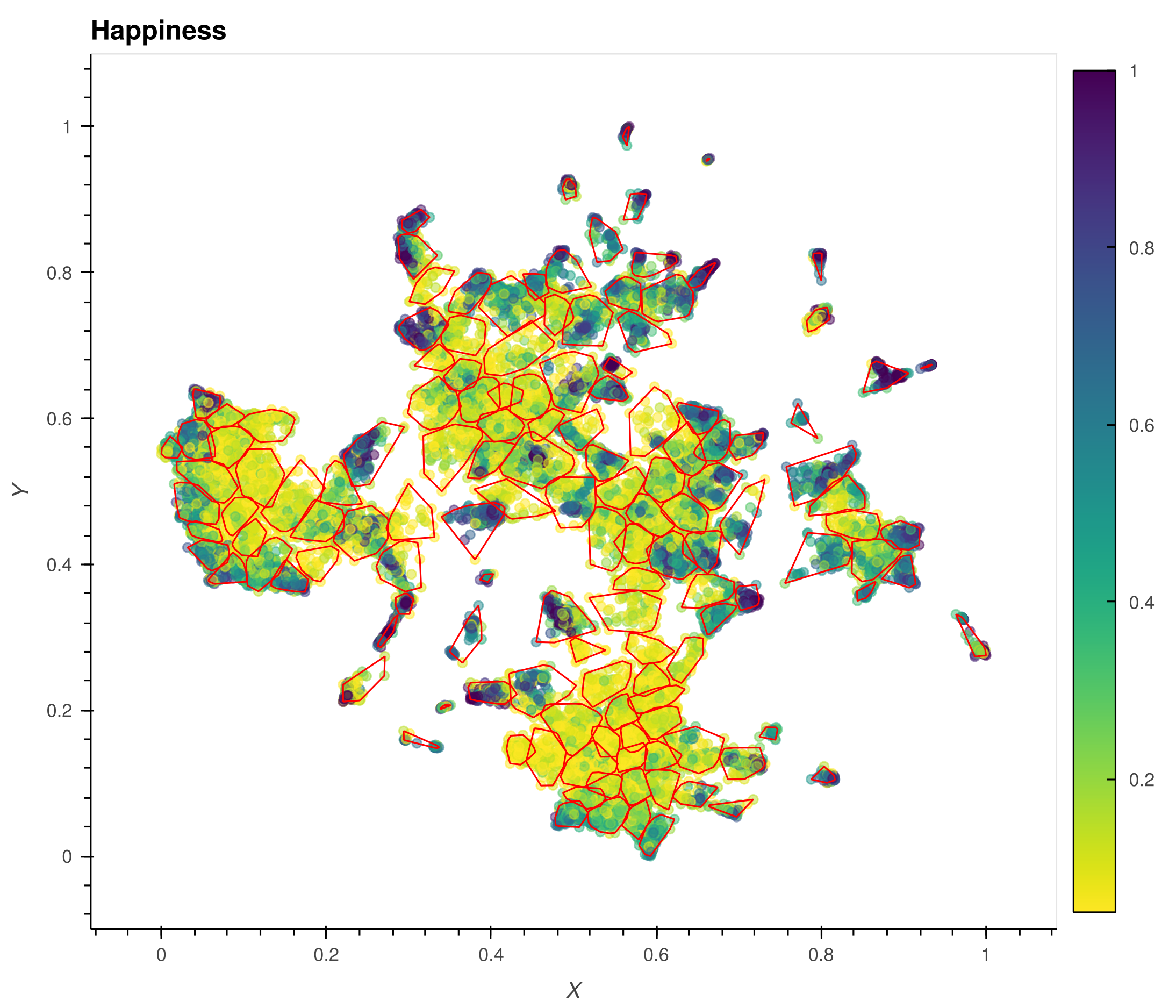
    \def\svgwidth{0.475\textwidth}
    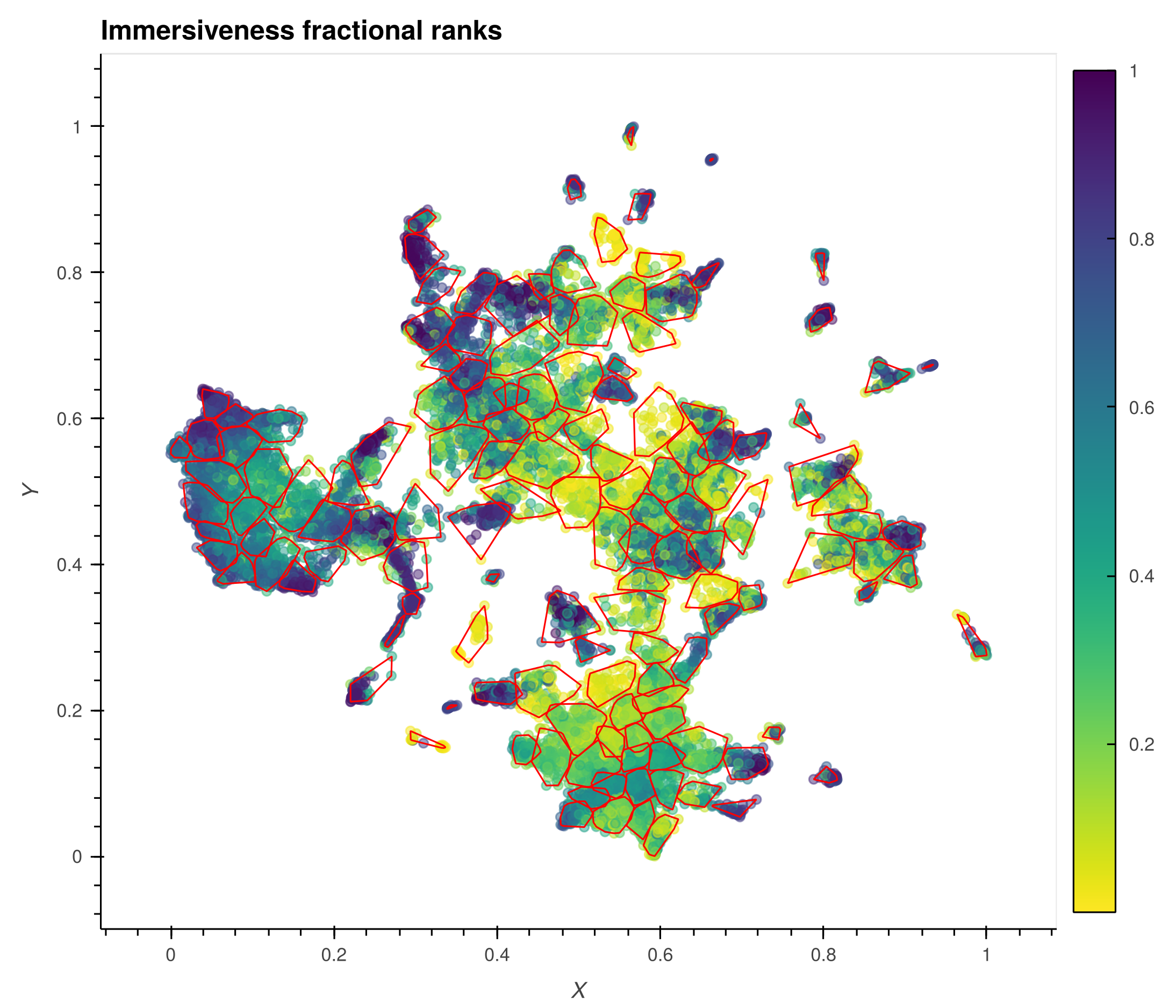
  \else
    \includegraphics[width=\halfwidth]{img/h_uma.png}
    \includegraphics[width=\halfwidth]{img/ifr_uma.png}
  \fi
    \caption{
Single-cell data exploration example.
The clustering UMAPS is colored by happiness and immersions to show their distributions respectively with convex hull borders of the clusters marked in red
    }
  \label{fig:hapimm_umaps}
\end{figure*}
From Fig.~\ref{fig:hapimm_umaps} it is clear that both the $h$ and $\iota$ metrics describe a similar underlying property of the data, namely how well it 
serves to describe covariational distances. In either case, denser regions correspond to more distinct attribution labels and higher happiness and 
immersiveness metrics. The difference between both metrics resides in their respective ranges as well as scaling with segmentation. 
The averages ($H, I$) are determined as a 
function of the underlying analyte parameters ($h,\iota$), but since both are dependent on the number of clusters in the segmentation ($M$) the scaling 
against the total amount of clusters can be expected to be different. In Fig.~\ref{fig:hapimm_umaps} we have chosen a cut through the distance hierarchy 
that yields $M=139$ clusters (for the single-cell data set). The choice of $M$ is not arbitrary, but note that only the full hierarchy preserves the 
distance- and topologic information of the full dataset. The value comes from the amount of clusters that will balance the amount of clusters against 
their mean sizes, see Fig.~\ref{fig:phase_transition}:~Bottom-Right~\cite{COPT}. Since the happiness values are more distinct than the 
individualized immersion 
values we have chosen to calculate and show the rank normalized immersions for visualisation purposes, in Fig.~\ref{fig:hapimm_umaps}. This shows that
the happiness and the immersion scores model a similar data property but that the quality of the property is not apparent for a given 
segmentation representation.
\begin{figure*}[t!]
 \centering
 \if\mybool1
    \graphicspath{{img}}
    \def\svgwidth{500pt}
    \input{img/Figure_happiness.pdf_tex}
 \else
    \includegraphics[width=500pt]{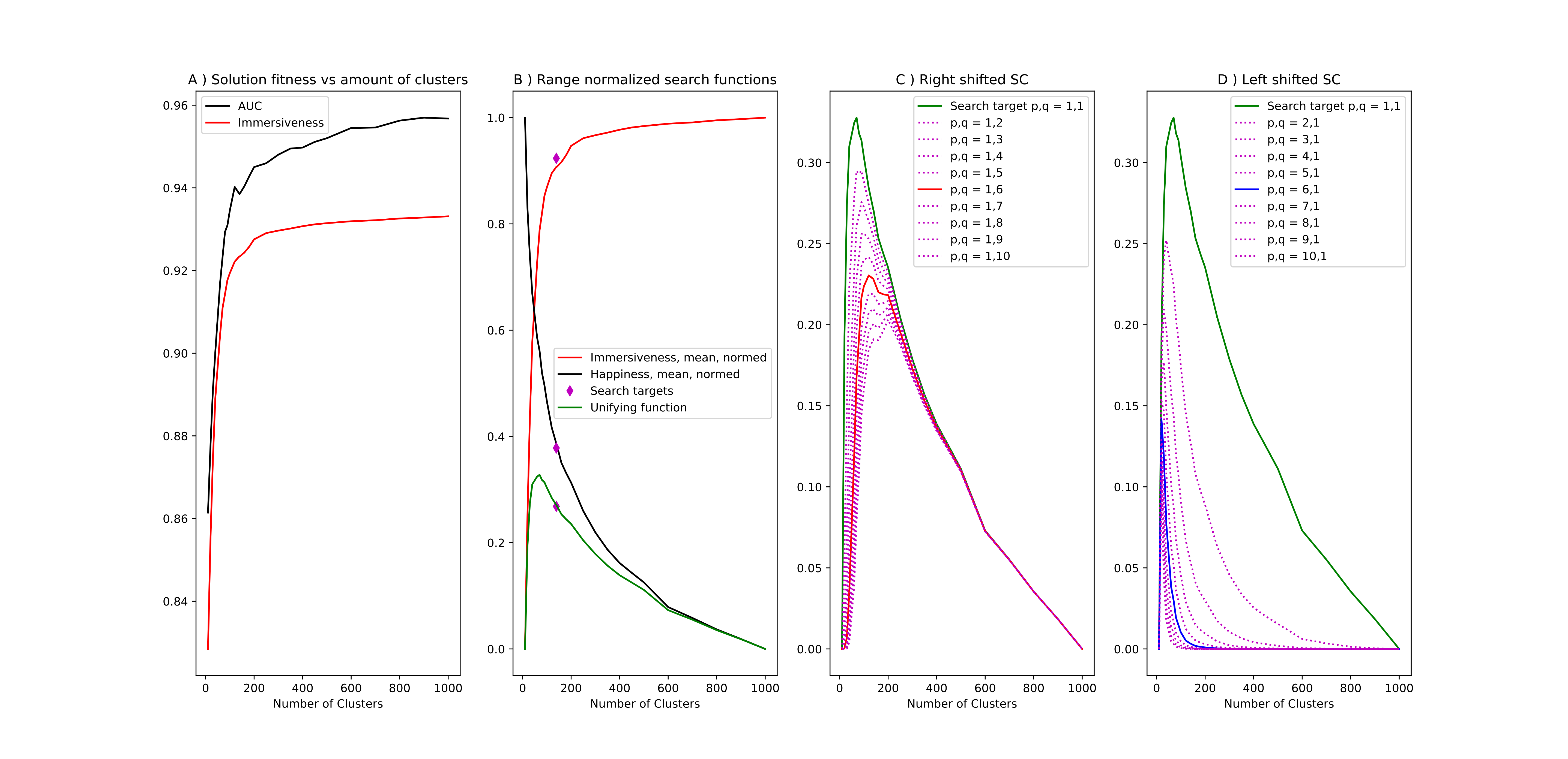}
 \fi
    \caption {
Single-cell data exploration example.
  A~) Showing the AUC and the immersiveness (average immersion) for the different clustering solutions.
  B~) Showing the range normalized mean happiness and mean immersion functions together with their unifying $\hat{G}$ function. The search targets are the 
chosen cut with $M=139$ clusters.
  C~) The right-shifted $\hat{G}$ functions under increasing powers of q.
  D~) The left shifted $\hat{G}$ function under increasing powers of p.
    }
 \label{fig:happiness_metrics}
\end{figure*}
In this section, we want to address the quality of the metrics. The immersiveness is expressed as the mean of all immersions and therefore 
does not treat all pairs on an equal footing during the rank assessment of similarity (or distance) values. 
This implies that the immersiveness is limited in resolution as compared to the full \ac{AUROC} calculation. 
In Fig.~\ref{fig:happiness_metrics}:~A we see that the same behaviour is captured by both metrics, but that the immersiveness value is lower and smoother 
as compared to the \ac{AUROC}. In Fig.~\ref{fig:happiness_metrics}:~B~) we can readily see the scaling behaviour of the two suggested metrics in relation 
to the number of clusters in the segmentation solution. The $H$ is monotonically decreasing with $M$ while the $I$ function is monotonically increasing. We 
can also see that the range normalized functions can be used to devise a unimodal search function but that the naive assumptions underlying such a 
formulation do not correspond to a high \ac{AUROC} score. We can however use the $p,q$ parameters to shift the peak to either emphasise happiness or 
immersiveness, see Fig.~\ref{fig:happiness_metrics}~C and Fig.~\ref{fig:happiness_metrics}~D.

Since the immersiveness is monotonically increasing but exhibits a stark derivative magnitude dependence versus the number of clusters one might expect a 
phase transition between the high and low \ac{AUC} cluster values. We can use the average of the derivatives of the immersions to obtain a better 
numerical sampling of the immersiveness derivative to capture this behaviour. The behaviour of the average of the immersion functions derivatives is 
depicted in Fig.~\ref{fig:phase_transition}. The transition point coalesces with a screening function that uses equation~(\ref{eq:geom}), with $p,q=1,1$, 
but offsets the number of clusters against the mean size of those clusters, as done in~\cite{COPT}. This tendency is depicted in 
Fig.~\ref{fig:phase_transition} 
\begin{figure*}[t!]
  \centering
    \centering
  \if\mybool1
    \graphicspath{{img}}
    \def\svgwidth{0.975\textwidth}
    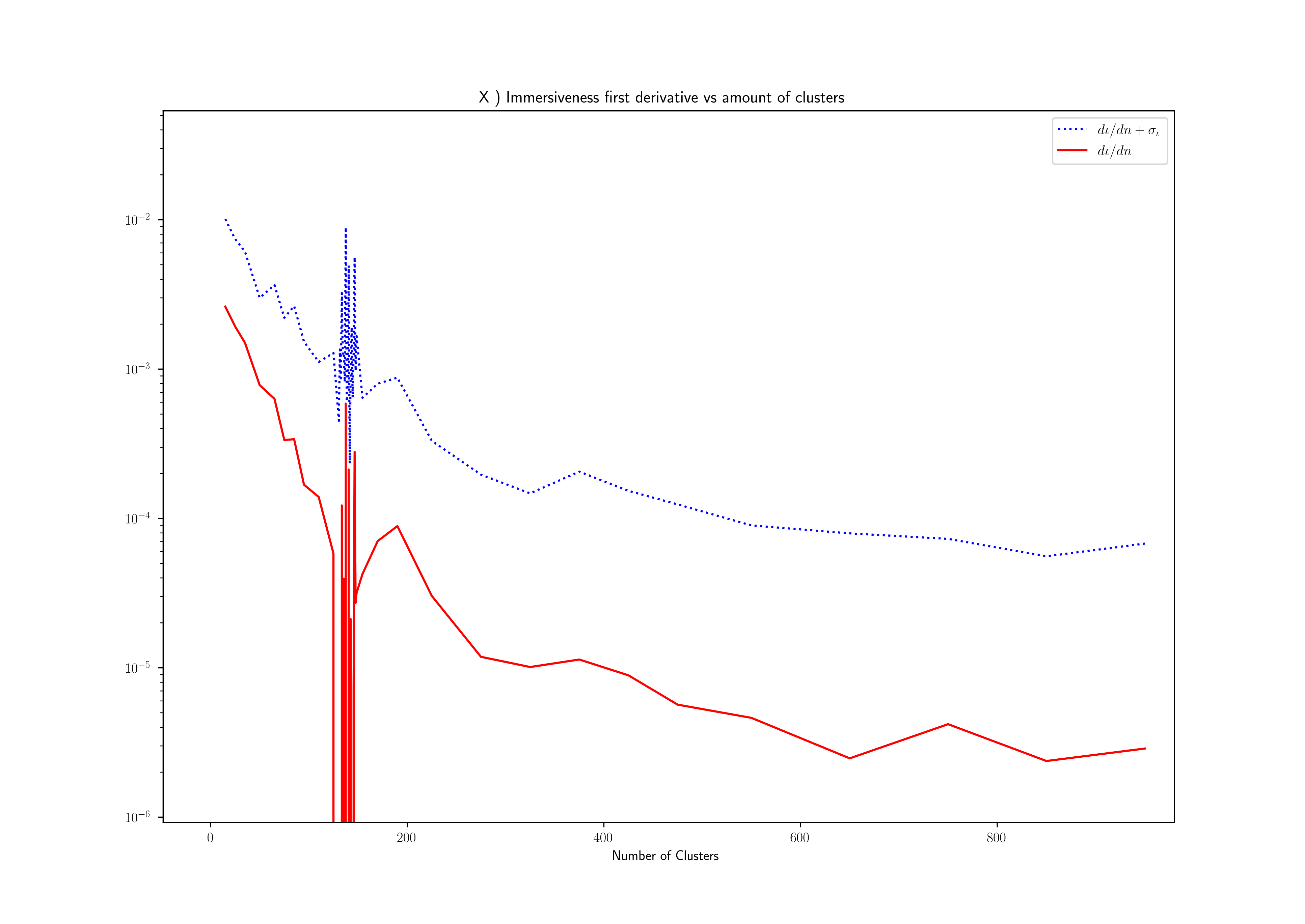
    \def\svgwidth{0.975\textwidth}
    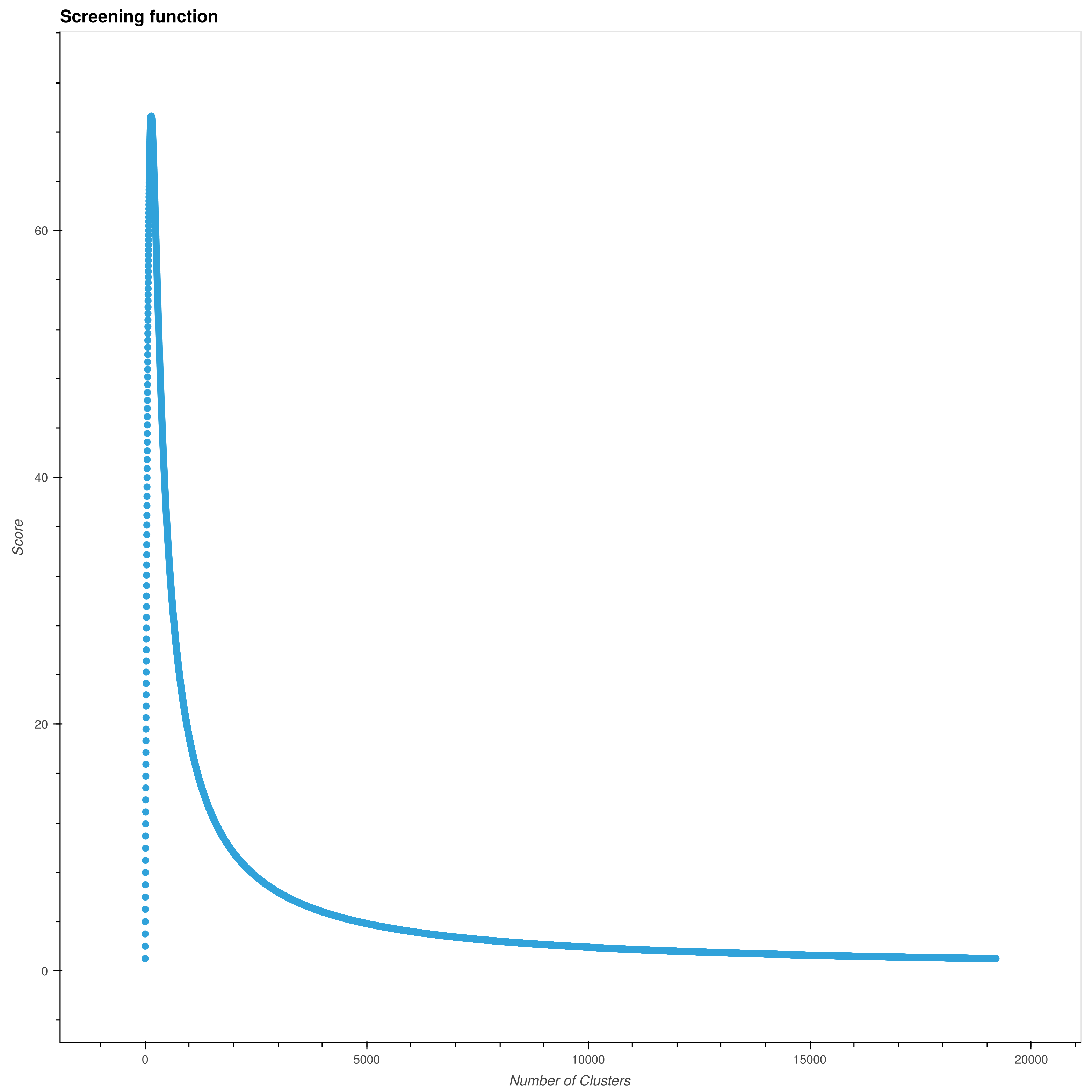
  \else
    {$\begin{matrix}
    \includegraphics[width=170pt]{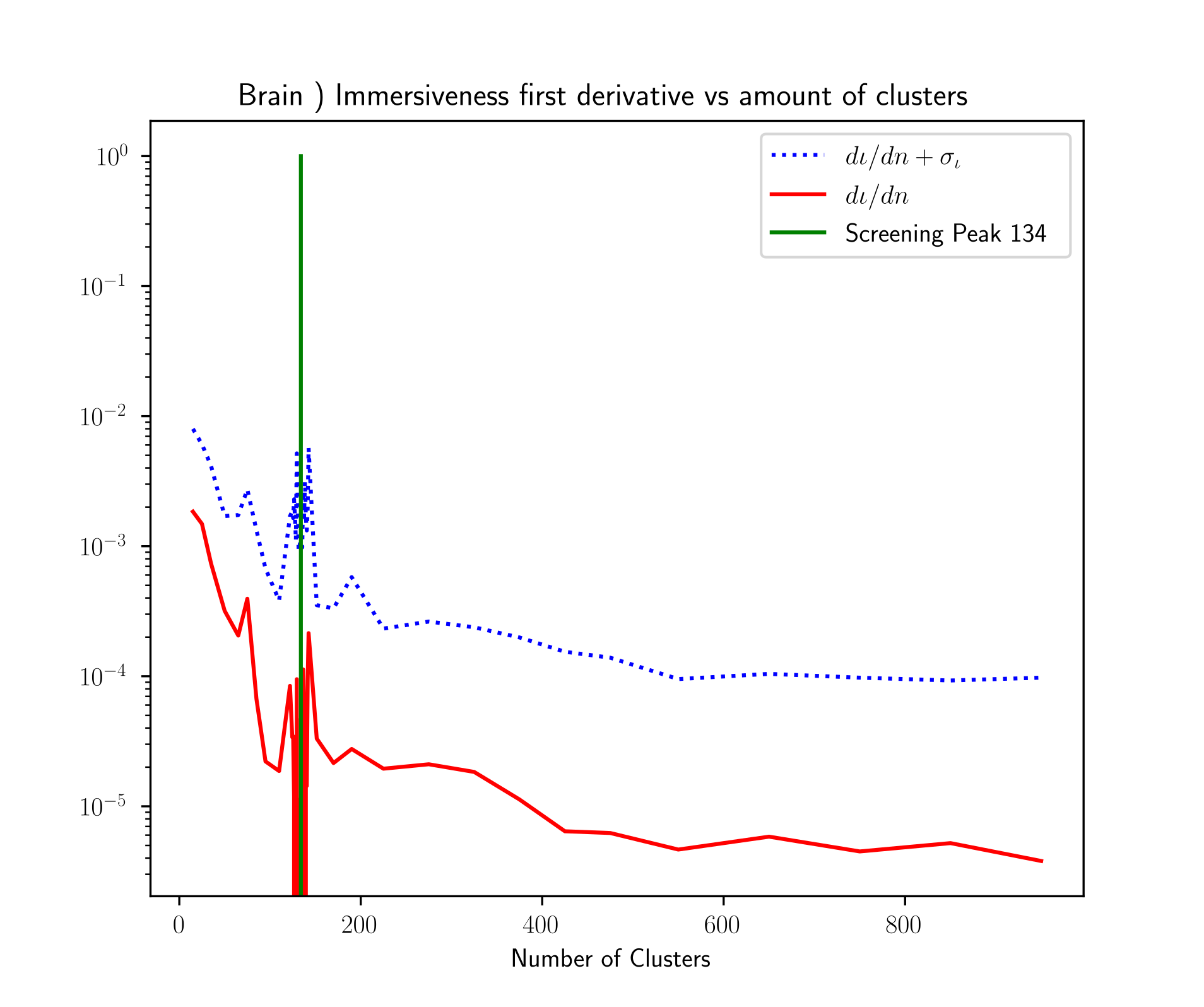} &
    \includegraphics[width=170pt]{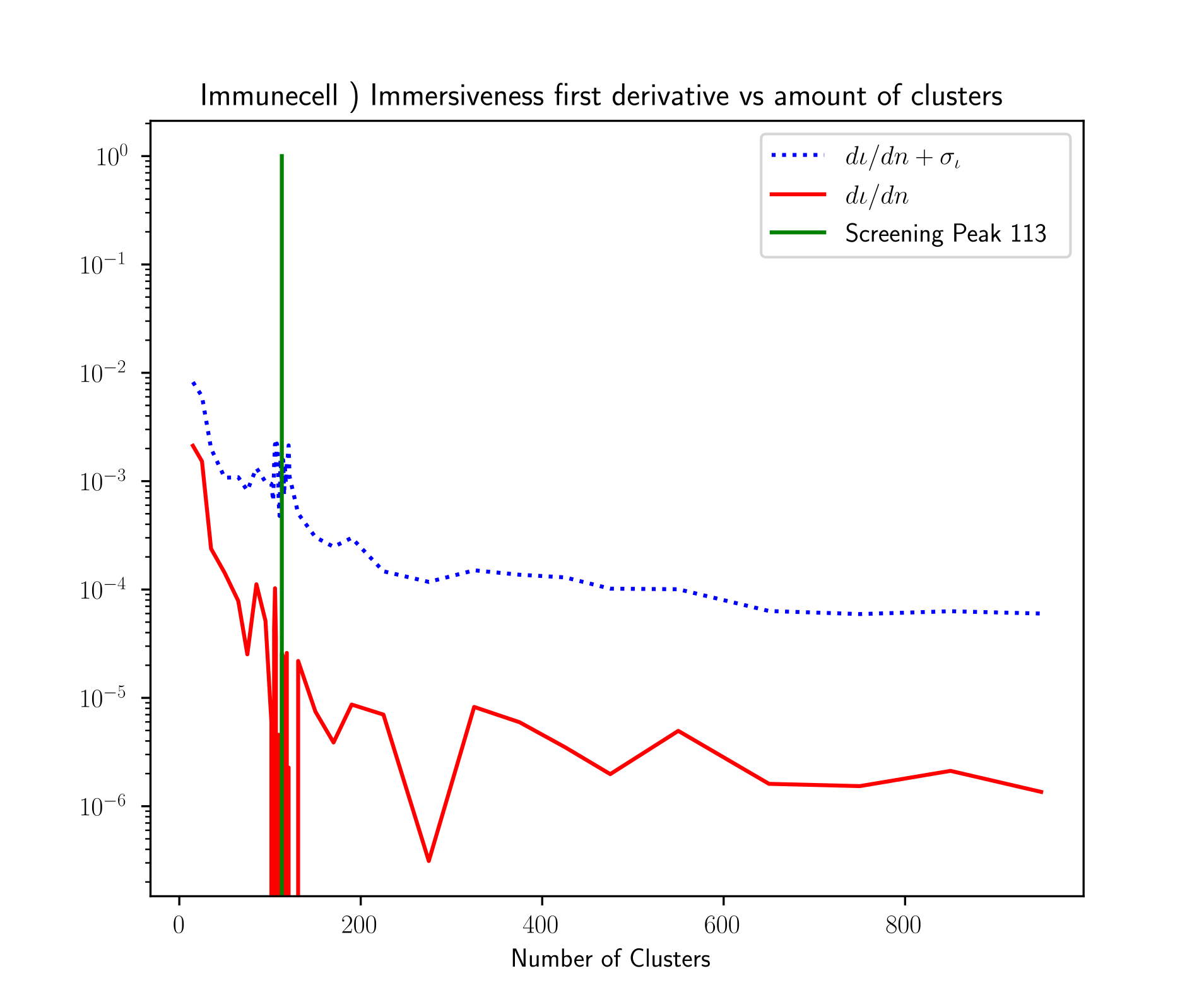} \\
    \includegraphics[width=170pt]{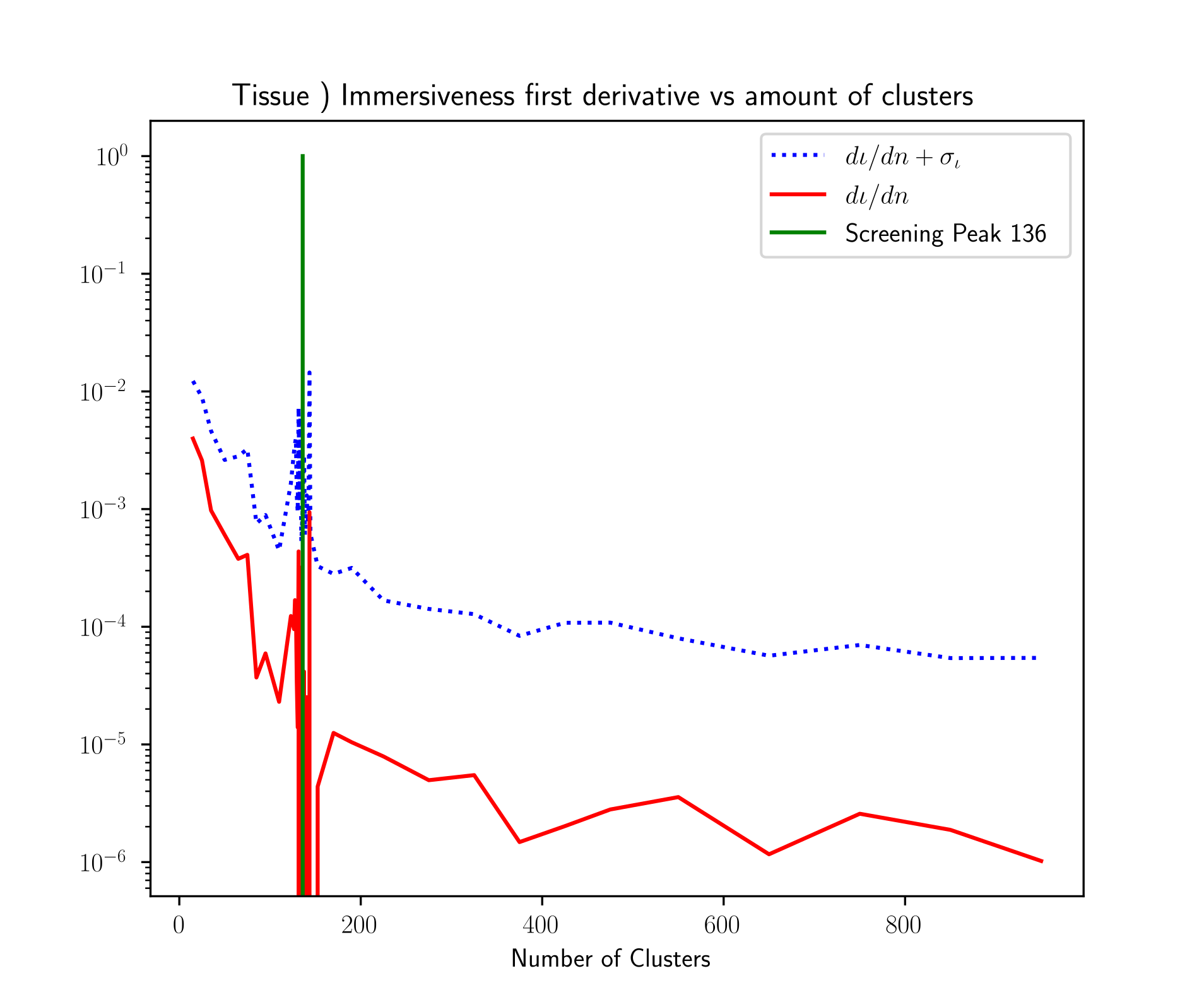} &
    \includegraphics[width=170pt]{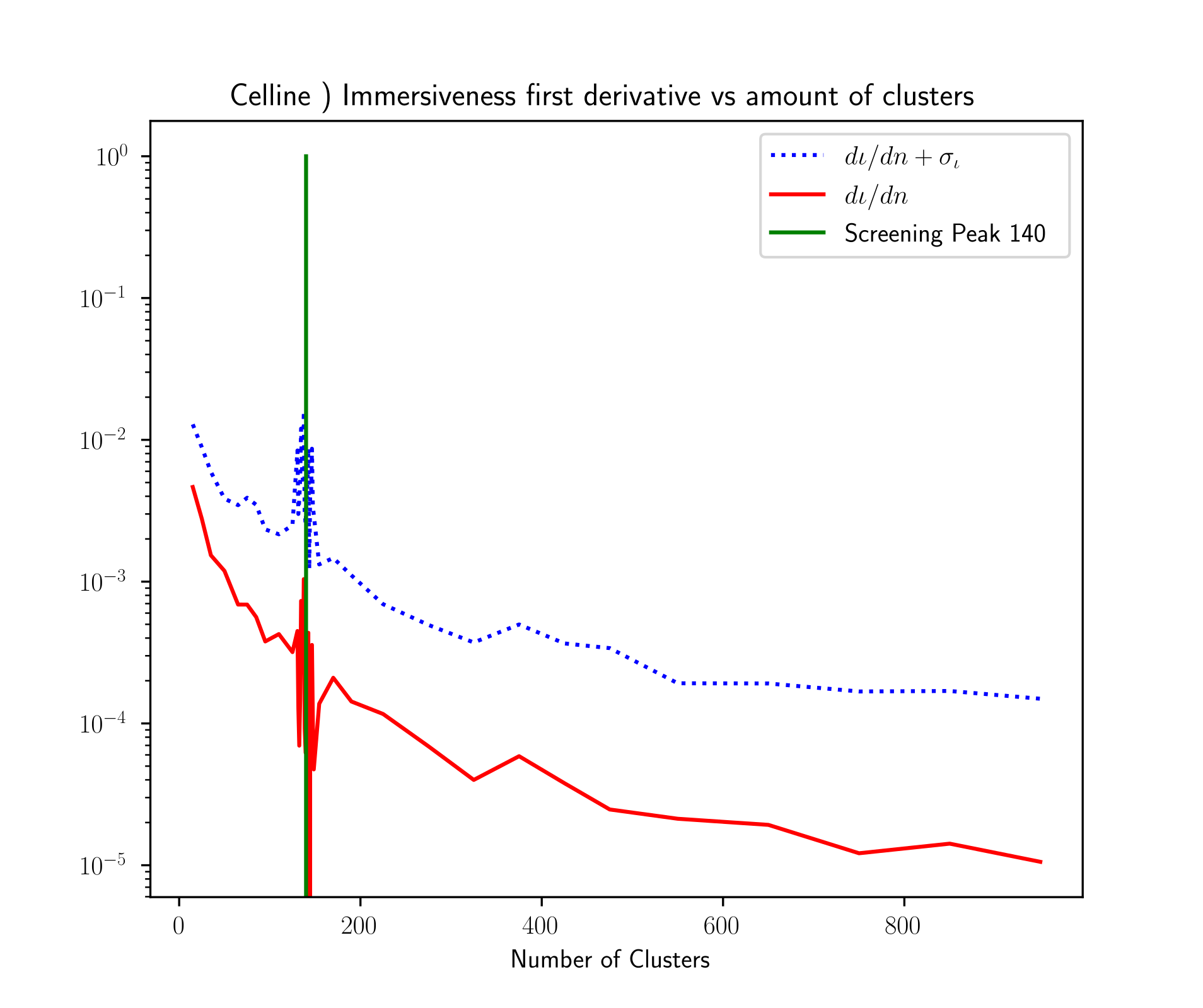} \\
    \includegraphics[width=170pt]{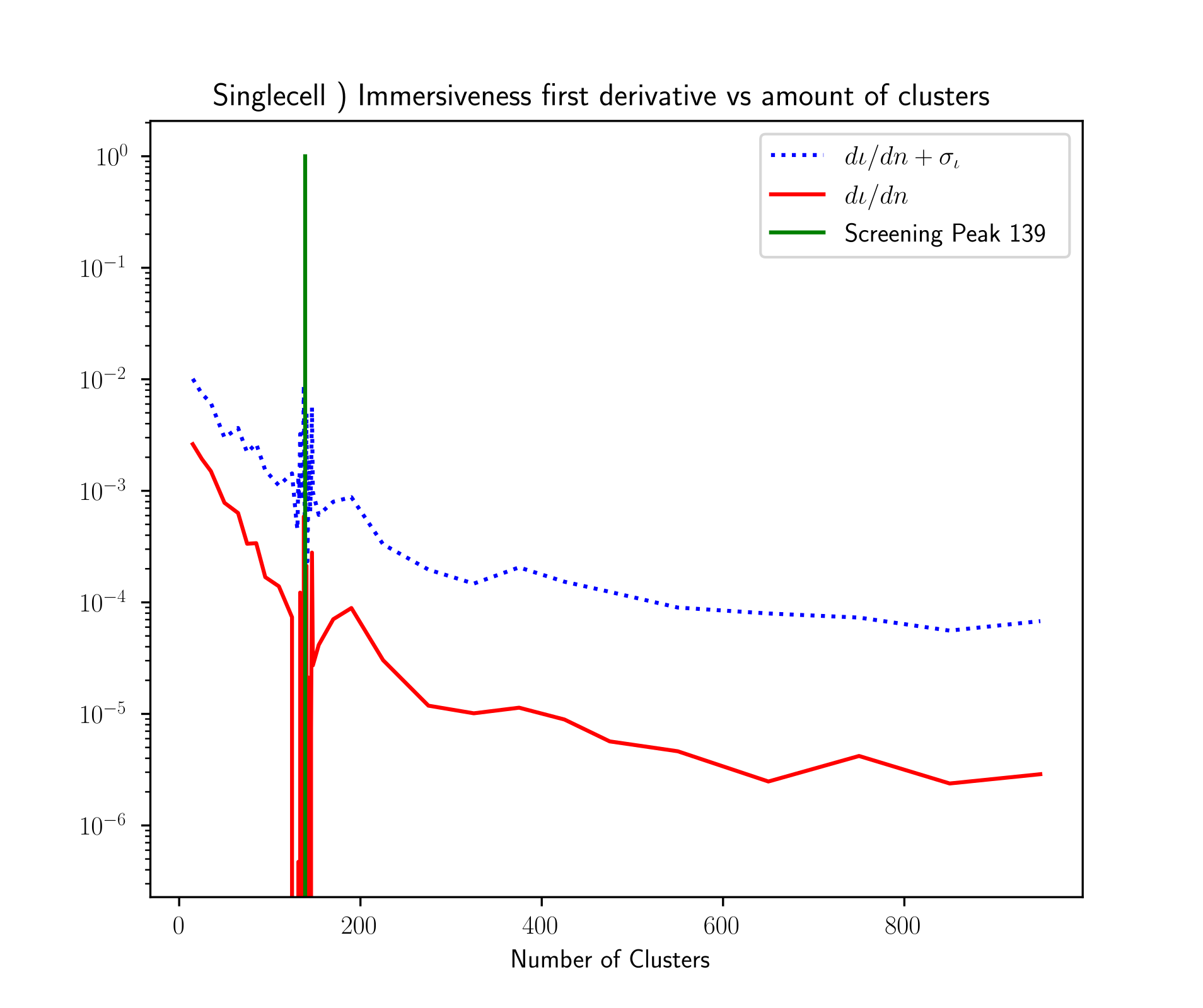} &
    \includegraphics[width=170pt]{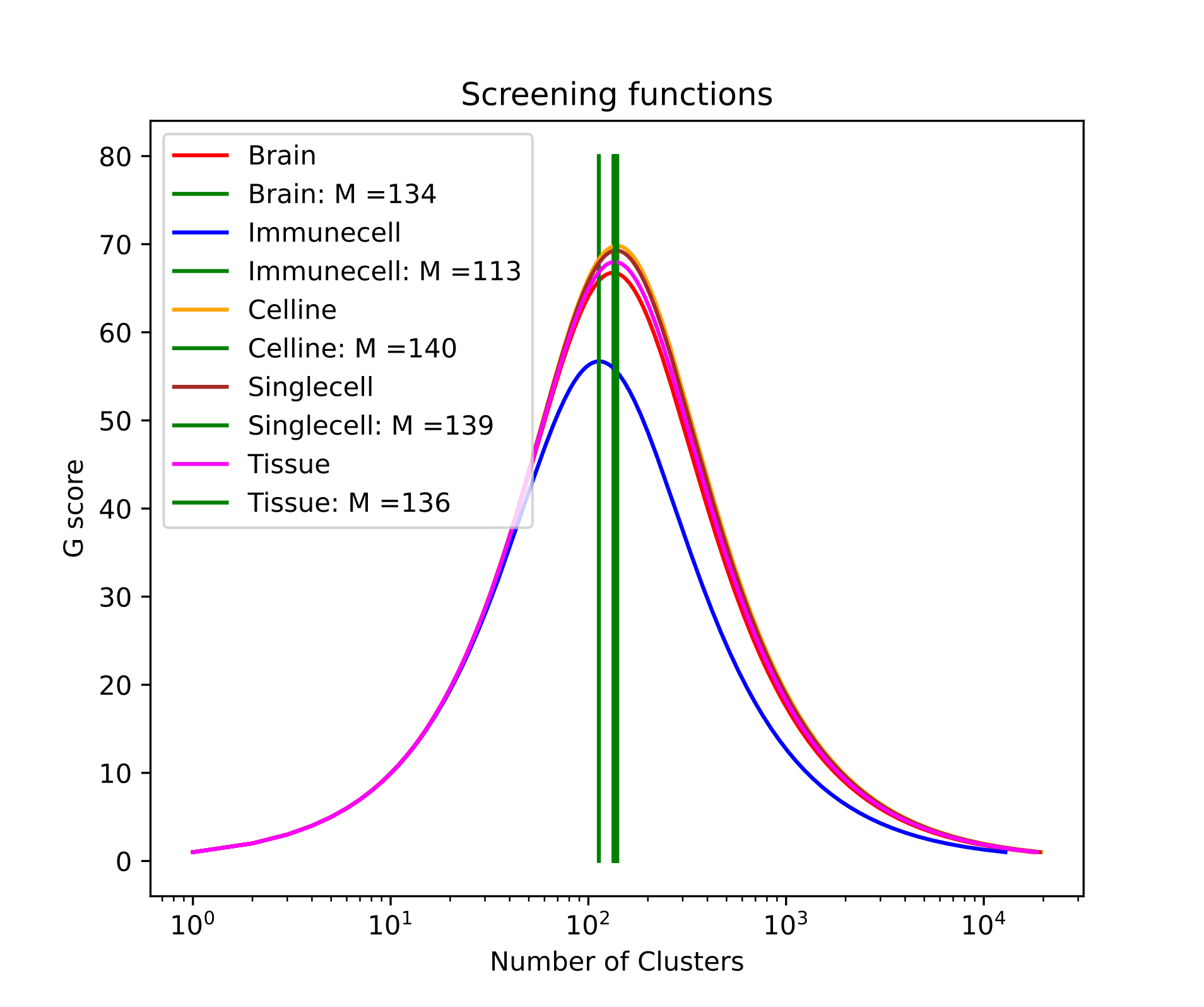} 
    \end{matrix}$}
  \fi
    \caption{
$X$~) The average of the immersion derivatives and the average with the first standard deviation added in ($X\in\{\text{Brain, Immune-cells, Single-cells, 
Cell-lines, Tissues}\}$). The largest derivative numerical instability occurs around $M=139$ clusters for the Single-cell dataset.
Screening functions~) The set of employed full unimodal, cluster size dependent, screening functions. These functions only assess the size of clusters and 
the number of clusters. Peak max value occurring for the Single-cell data at $M=139$ clusters.
    }
  \label{fig:phase_transition}
\end{figure*}
where we have assessed how the average derivative of the immersion values scales with respect to the number of clusters. It is clear from 
Fig.~\ref{fig:phase_transition} that there is an abrupt change in the individualized analyte \ac{AUC} values and that this peak position coalesces with 
one found by using a unimodal search function that offsets the size of clusters versus the number of clusters in the segmentation cut through an 
agglomerative hierarchical clustering solution that employs Ward's partitioning strategy. We can see in Fig.~\ref{fig:happiness_metrics} that both 
the \ac{AUC} and immersiveness values are monotonically increasing functions but that the gain over increasing cluster numbers drops at higher cluster 
numbers. These two regimes, where the gain is large versus small, are well separated by both the numerical behaviour of the average immersion derivatives 
as well as the size-dependent $G$ function. To conclude we can see that the direct numerical derivative disruption of the immersiveness follows the position
of the information capacity as well as the unimodal size-dependent search function. We can observe that the immersiveness and happiness metrics 
drive segregation and aggregation respectively. This observation is aligned with our theoretical understanding of the metrics.

\subsection*{ Composition }

\begin{figure*}[t!]
 \centering
    \centering
 \if\mybool1
   \graphicspath{{img}}
   \def\svgwidth{0.475\textwidth}
   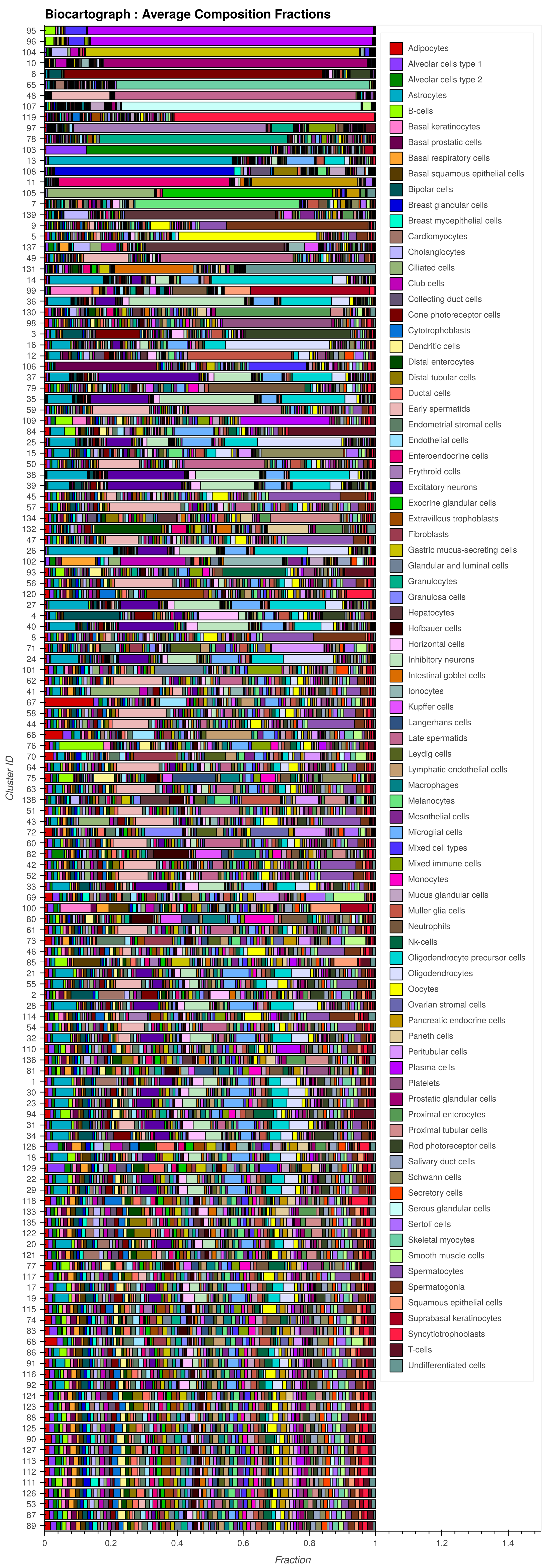
   \def\svgwidth{0.475\textwidth}
   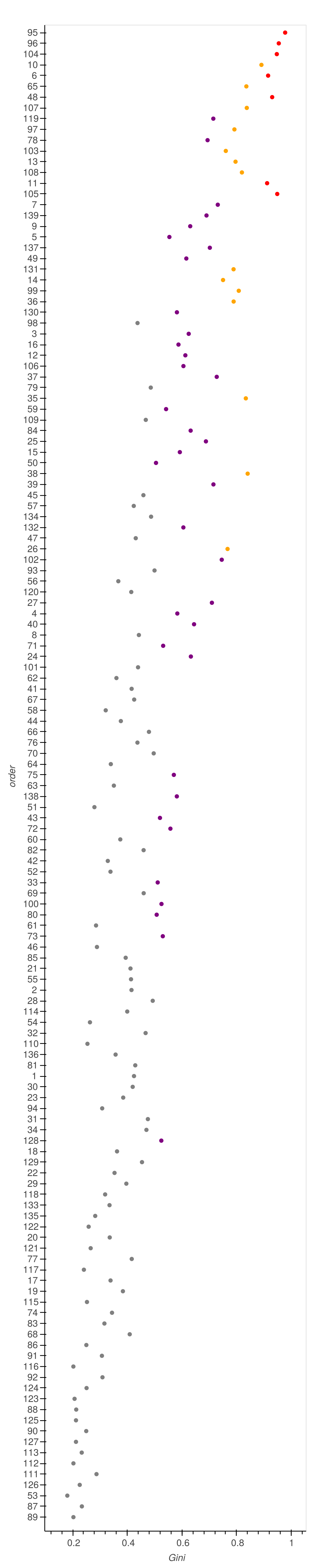
 \else
    \includegraphics[width=\halfwidth]{img/bioc_clu_frac_com.png}
    \includegraphics[width=100pt]{img/cluster_specificit.png}
 \fi
    \caption{
Single-cell data exploration example.
Left~) Cluster average compositions. Right~) Cluster Gini coefficients
    }
 \label{fig:analyte_cluster_composition}
\end{figure*}

In this section, we want to address whether the stated relationship between the clustering \ac{AUROC} measure and the compositional Gini coefficient holds
for our labelled data. Furthermore, the compositional dependence can be explored for the different data sets to increase our understanding
of the connections between composition and covariation.
\begin{figure*}[t!]
 \centering
    \centering
 \if\mybool1
   \graphicspath{{img}}
   \def\svgwidth{0.475\textwidth}
   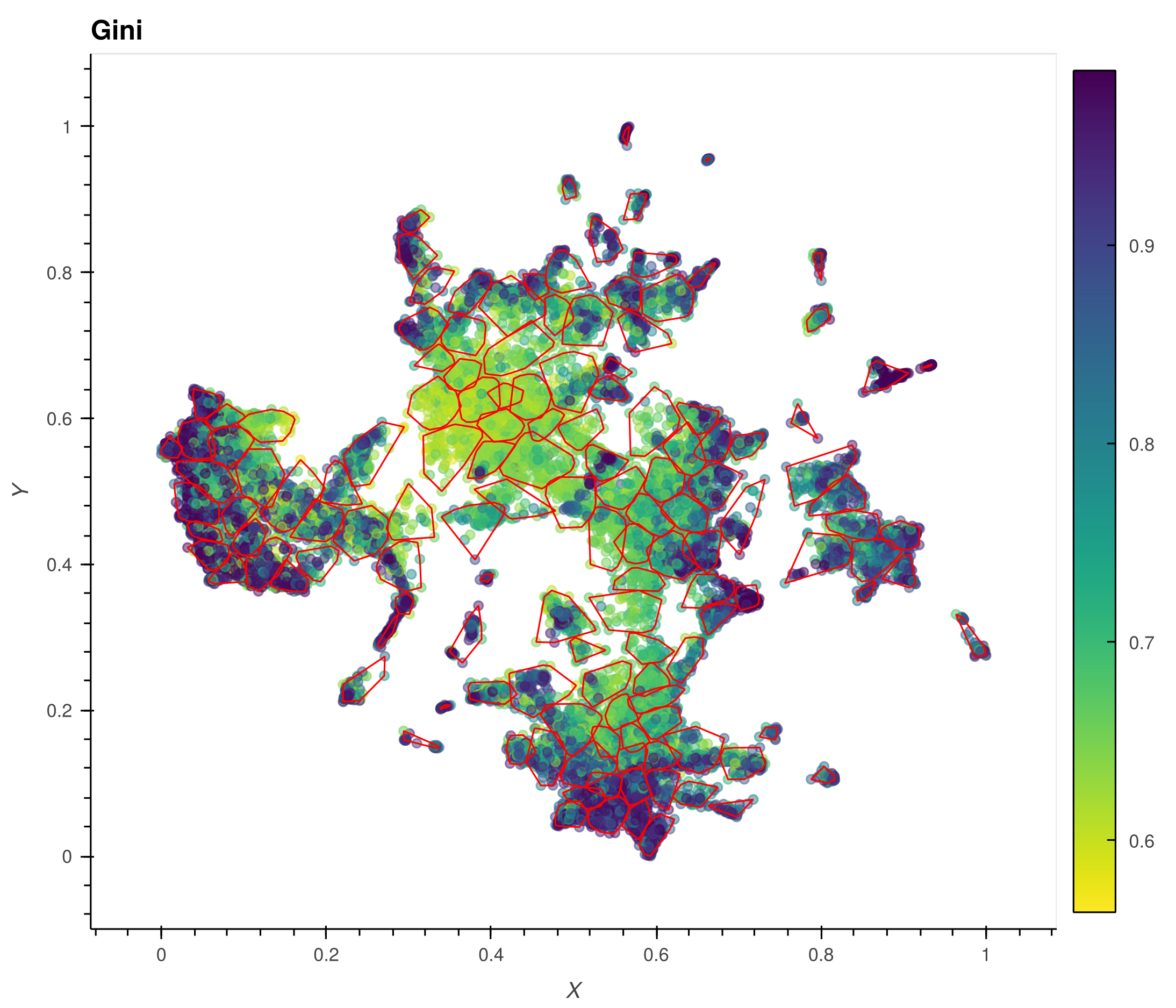
   \def\svgwidth{0.475\textwidth}
   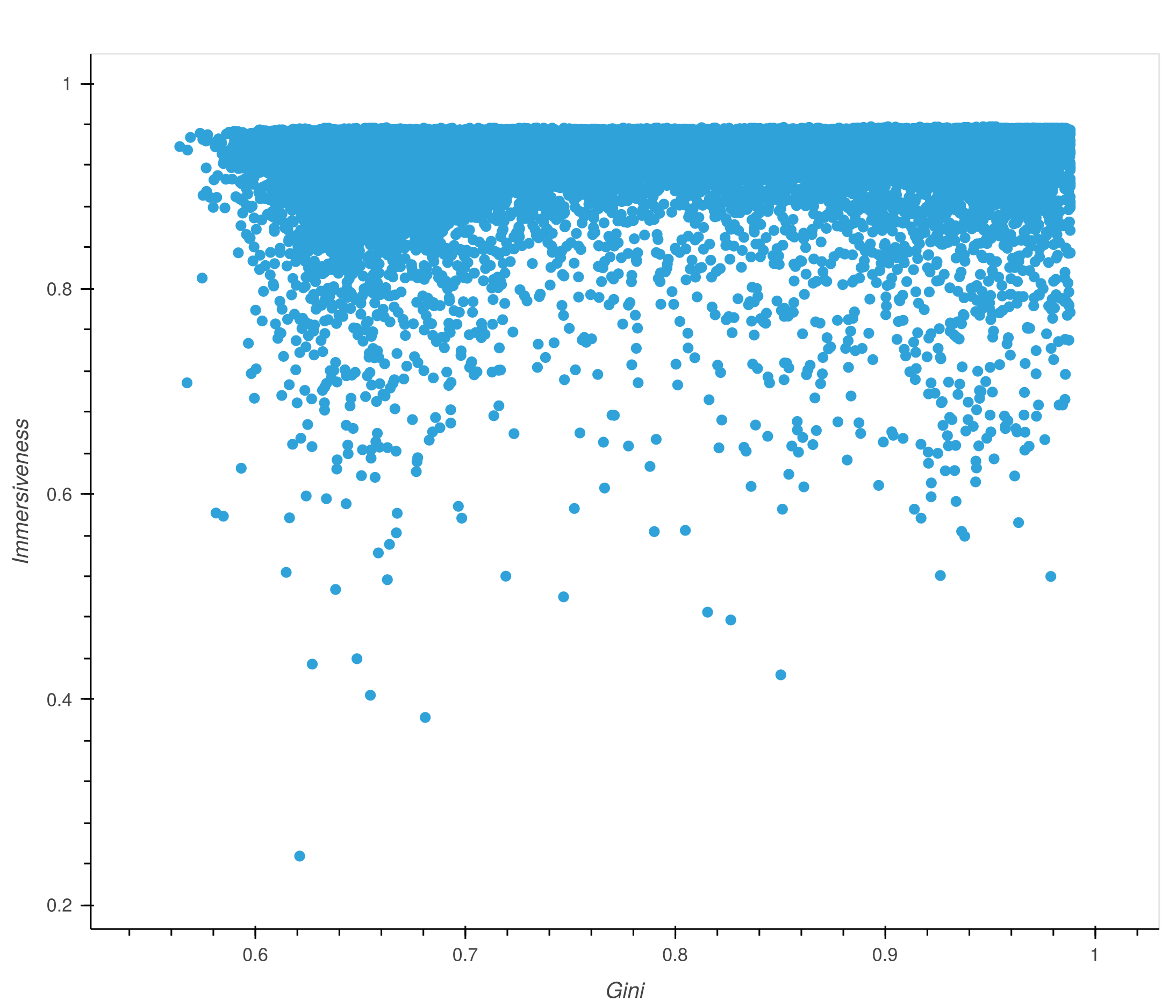
 \else
    \includegraphics[width=\halfwidth]{img/Gini_uma.png}
    \includegraphics[width=\halfwidth]{img/i_gini_s.png}
    \includegraphics[width=300pt]{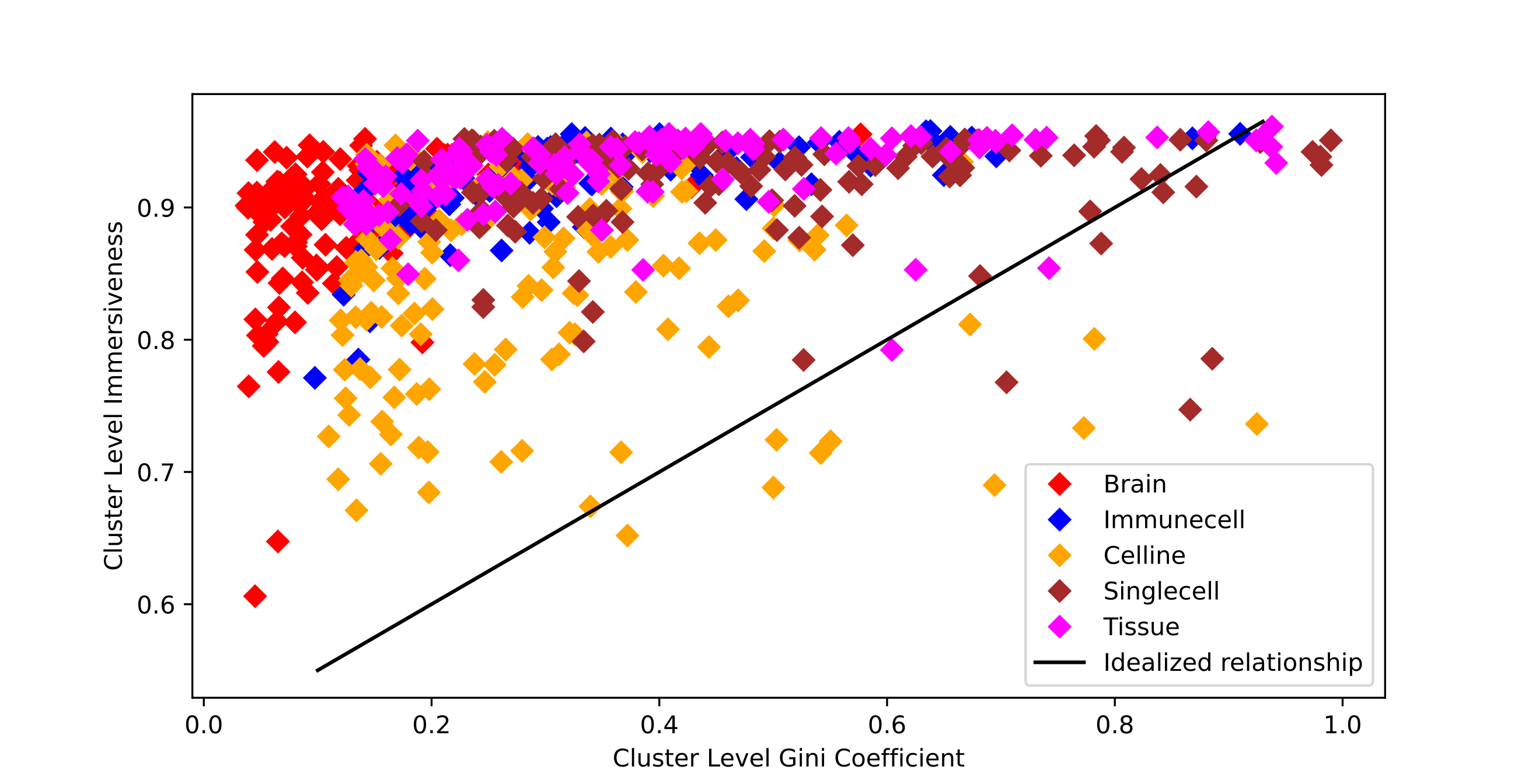}
 \fi
    \caption {
Single-cell data exploration example.
        Left~) The Gini coefficient of each analyte's absolute composition in the input.
        Right~) The Gini coefficient versus immersion shows no direct relationship.
        Bottom~) The Cluster level aggregated comparisons between Immeriveness values and their respective Gini coefficients for the five 
data-set screened peak max position segmentation solutions.
    }
 \label{fig:gini_umap}
\end{figure*}
In Fig.~\ref{fig:gini_umap} we can see that there is no strong direct relationship between the absolute compositional 
coefficients~\cite{composition_benchmark_2016} and the 
immersion values. We can however see that higher compositional coefficients, on average, correspond to higher immersion values. This weak 
dependence could be understood in that the covariational distances are relative metrics calculated for a clustering solution. All the 
compositional coefficients are calculated atomically for each label grouping per analyte without dependence on the segmentation so the result 
is expected. From Fig.~\ref{fig:gini_umap} we find that many of the clusters analytes belonging to clusters with a high gini coefficient contain 
analytes that are close in absolute compositions label profiles. The clusters with the highest gini coefficients ($\approx 5\%$ of clusters) have 
absolute, fractional and sum-normed compositions that are similar in group label profiles (sample qualities). This disproves the notion that clusters
with high immersion scores also obtain high Gini coefficients. This was explored and found to be a consistent observation for all datasets and
we show the typical coefficient fall-off behaviour and compositional make-up in Fig.~\ref{fig:analyte_cluster_composition}.
For each cluster, we have chosen to calculate the typical compositional character of each cluster. The depiction in 
Fig.~\ref{fig:analyte_cluster_composition} illustrates what the compositional profile of a randomly drawn analyte would look like. The fractions thereby 
do not relate directly to an absolute number of label types for each cluster but rather what the most common expression distribution for any analyte 
belonging to that cluster would look like.

\section{ Discussion } \label{sec:discuss}

The main finding of this article is the transition state behaviour of the immersiveness metric that is intimately related to the traditional \ac{AUROC} of
the clustering solution. We will elaborate on some of the details in this section.

Since the sample labels are not necessarily strongly linked with the underlying covariation one should not expect a strong relationship between the Gini 
coefficient and the immersion values for a general clustering approach on labelled data. This is evident as being the case in our single-cell 
data, see Fig.~\ref{fig:gini_umap}~(~Right~) as well as Fig.~\ref{fig:analyte_cluster_composition}.

The $\iota$ metric is costly to calculate for any segmentation step in the family of segmentations belonging to the full hierarchy. Generally, algorithmic 
scaling is similar to or worse for the immersion calculation than the efficient hierarchical construction. Furthermore, since numerical derivatives are inherently noisy, 
we could not use the average $d\iota/dn$ directly to efficiently construct a search function concerning \ac{AUC} completeness during segmentation 
construction, however, the $\sigma^2_\iota$ works well for this purpose. The disruption of the immersion derivatives coalescence with the 
cluster size screening implies that it might be a better candidate for 
direct optimisation strategies~\cite{COPT}. We note that the immersiveness calculation is more efficient by employing our formalism than calculating
related measures such as the Davies-Bouldin score~\cite{daviesbouldin1979} or the Calinski-Harabasz index~\cite{calinskiharabasz1974}.

We note that the observed transition only applies to the type of segmentation strategy where the clustering is formed in a systematically connected fashion.
Where the search function is guaranteed analytic by the agglomerative hierarchical approach. This is not necessarily the case for other popular clustering 
approaches~\cite{traag_louvain_2019}. The coalescence of the numerical disruption of the immersiveness derivatives and the size-dependent $G$ function is 
expected to hold for other agglomerative hierarchical approaches than Wards, as explored in the supplementary information. The $\iota$ transition property 
was explored for both real-world and simulated data using several \ac{AHC} approaches and is shown to work as an assessment for 
hierarchical clustering solutions.

The immersiveness or immersion score lets us decide on cluster numbers using the covariation structure of the data. A comparison with other internal 
metrics exists in the supplementary where compositionally inferred labels are compared to cluster segmentation labels. We could not use label set scores to 
decide on cluster numbers in a fashion directly connected to the covariational
structure. Metrics such as mutual information and ARI score etc are strictly dependent on comparing label sets and not those labels' ability to 
model covariation thereby rendering them useless in our context. The Davies-Bouldin score and Calinski-Harabasz index, but not the Silhouette score or 
Dunn index, are expected to exhibit a similar transition state as the immersiveness evaluated here. The comparative study of the transition behaviour 
of those metrics could form the basis for a future study.

\section{ Conclusions }

Comparing Fig.~\ref{fig:gini_umap}, Fig.~\ref{fig:hapimm_umaps} and Fig.~\ref{fig:analyte_cluster_composition} it is clear that the Gini coefficient 
describes the typical cluster composition well but cannot generally be used to re-create the \ac{AUC} value belonging to the cluster group analytes.

For our clustering approach, we can find a transition point of the \ac{AUC} that relates to the number of clusters we need to describe the data well. We 
see that this behaviour occurs in both synthetic as well as real-world data using a plethora of \ac{AHC} linkage approaches. This has a bearing on the 
modelling of biological function since the minimal set of uniquely covarying groups of analytes forms a basis for solving
the encoding problem while discarding tightly covarying contributions during solution construction. We can show that the coarse graning transition point 
coalesces with the strategy of finding a size distribution offset between cluster content and amount of clusters. Thereby potentially facilitating a 
faster and more efficient strategy for finding a meaningful representation of the data under study.

\newpage

\bibliography{metclust.bib}

\newpage

\section{Supplementary information}

\begin{figure*}[t!]
 \centering
    \centering
 \if\mybool1
   \graphicspath{{img}}
   \def\svgwidth{0.65\textwidth}
   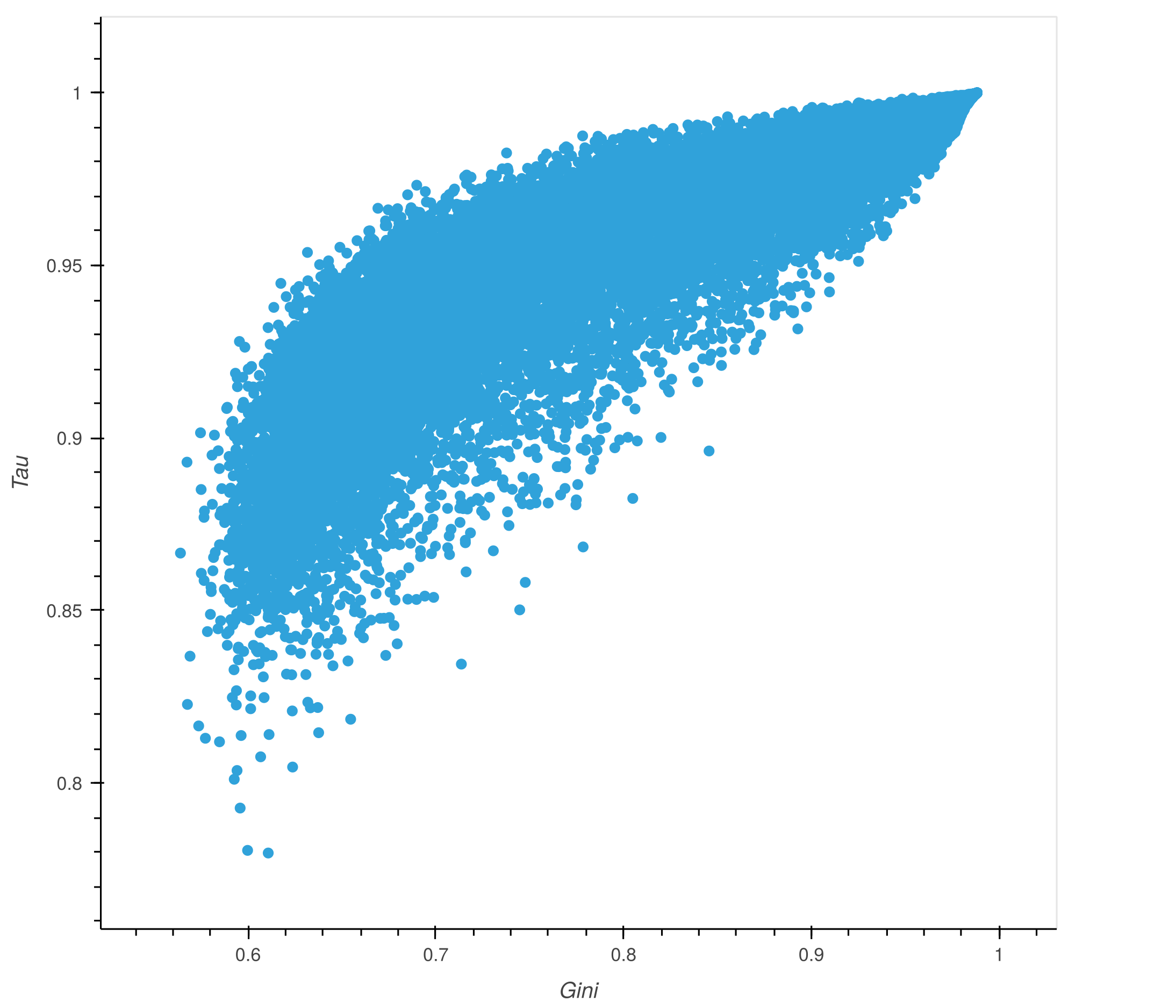
 \else
    \includegraphics[width=250pt]{img/TauGini_s.png}
 \fi
    \caption {
Single-cell data exploration example.
        The relationship between the (absolute) compositional Gini specificities and Tau specificities
    }
 \label{fig:tau_gini}
\end{figure*}

\subsection*{Software}

\subsubsection*{Environment}

\href{https://github.com/rictjo/versioned-nix-environments/blob/main/env/versioned_R_and_Python.nix}{versioned R and Python nix environment definitions}

\vspace{0.2cm}
Enter the environment using the \href{https://nixos.org}{nix package} manager:
\begin{lstlisting}
$ nix-shell versioned_R_and_Python.nix
\end{lstlisting}

\subsubsection*{Main package}
\url{https://github.com/rictjo/biocarta}

or

\url{https://github.com/richardtjornhammar/biocartograph}

\subsection*{Installing}
If you are not using nix then install the $Biocartograph$ software using pip:
\begin{lstlisting}
$ pip install biocartograph
\end{lstlisting}

\subsubsection*{The Biocartograph Single-cell Solution}

\href{https://rictjo.github.io/?https://gist.githubusercontent.com/rictjo/d5327e3a85bca48a75dbe119fe98bf5b/raw/839a87e79207abc246fa0287c26d2e42dee4fdda/index.html}{Interactive Clustering Graphs}

\subsubsection*{The Immersiveness, Happiness and Gini Cluster configuration}

\href{https://rictjo.github.io/?https://gist.githubusercontent.com/rictjo/a50c7e5dfbe08bbf23b1d6730624ff2b/raw/2cc2aa7132343a14ffdb421c893e263b487a37ad/index.html}{Graphs of Coefficients}

\subsubsection*{Immersion derivatives}
\href{https://gist.github.com/rictjo/9d03cb970428aa44360ab89e40cbc0f2}{Single cell derivatives}

\href{https://gist.github.com/rictjo/fefdacfe7314cfbd632753b7468860ec}{All data-set derivatives}

\href{https://gist.github.com/rictjo/ee16f6da2e96962873c6ba16bc279649}{Screening functions}
%
%
\href{https://gist.github.com/rictjo/44e09a05e020d92ab3a296118878b581}{Immersiveness (AUC + error estimates), several linkage methods}

\href{https://gist.github.com/rictjo/87586c02593164af8d5af910a02314fc}{Information Capacity, several linkage methods}

\href{https://gist.github.com/rictjo/7c34f4e2ce77c49e75526f31915fb62c}{Toy data generation example}

\subsubsection*{Cluster Composition Gini coefficients}

\href{https://gist.github.com/rictjo/2e5419f2c27718e3ac61145735410d38}{Specificity values of single cell clusters}

\href{https://gist.github.com/rictjo/d9002d0e26fe13fc51f64c9fc69e45d4}{Cluster level Gini coefficients}

\subsubsection*{Data availablity}
Data can be obtained upon reasonable request to the author

\end{document}